\documentclass[journal=jacsat,manuscript=article]{achemso}

\usepackage{chemformula} 
\usepackage[T1]{fontenc} 
\usepackage{caption}
\usepackage{subcaption}
\usepackage[leftcaption]{sidecap}

\usepackage{soul}
\usepackage{newfloat}
\usepackage{hyperref}
\usepackage{amssymb}

\usepackage{tikz,xcolor,hyperref}
\usepackage{graphicx}

\definecolor{lime}{HTML}{A6CE39}
\DeclareRobustCommand{\orcidicon}{%
	\begin{tikzpicture}
	\draw[lime, fill=lime] (0,0) 
	circle [radius=0.16] 
	node[white] {{\fontfamily{qag}\selectfont \tiny ID}};
	\draw[white, fill=white] (-0.0625,0.095) 
	circle [radius=0.007];
	\end{tikzpicture}
	\hspace{-2mm}
}

\foreach \x in {A, ..., Z}{%
	\expandafter\xdef\csname orcid\x\endcsname{\noexpand\href{https://orcid.org/\csname orcidauthor\x\endcsname}{\noexpand\orcidicon}}
}

\author{Theodoros D. Bouloumis} 
\affiliation{Okinawa Institute of Science and Technology Graduate University, Onna, Okinawa 904-0495, Japan}
\email{Theodoros.Bouloumis@oist.jp}
\author{Domna G. Kotsifaki}
\affiliation{Okinawa Institute of Science and Technology Graduate University, Onna, Okinawa 904-0495, Japan}
\altaffiliation{Natural and Applied Sciences, Duke Kunshan University, No. 8 Duke Avenue, Kunshan, Jiangsu Province, 215316, China}
\author{S\'{i}le {Nic Chormaic}} 
\affiliation{Okinawa Institute of Science and Technology Graduate University, Onna, Okinawa 904-0495, Japan}

\title[An \textsf{achemso} demo]
  {Enabling self-induced back-action trapping of gold nanoparticles in metamaterial plasmonic tweezers 
}

\abbreviations{LSP, FIB, SEM, MSP}
\keywords{American Chemical Society, \LaTeX}

\begin{document}

\begin{tocentry}

    \centering
    \includegraphics[width = 70 mm, height = 45 mm]{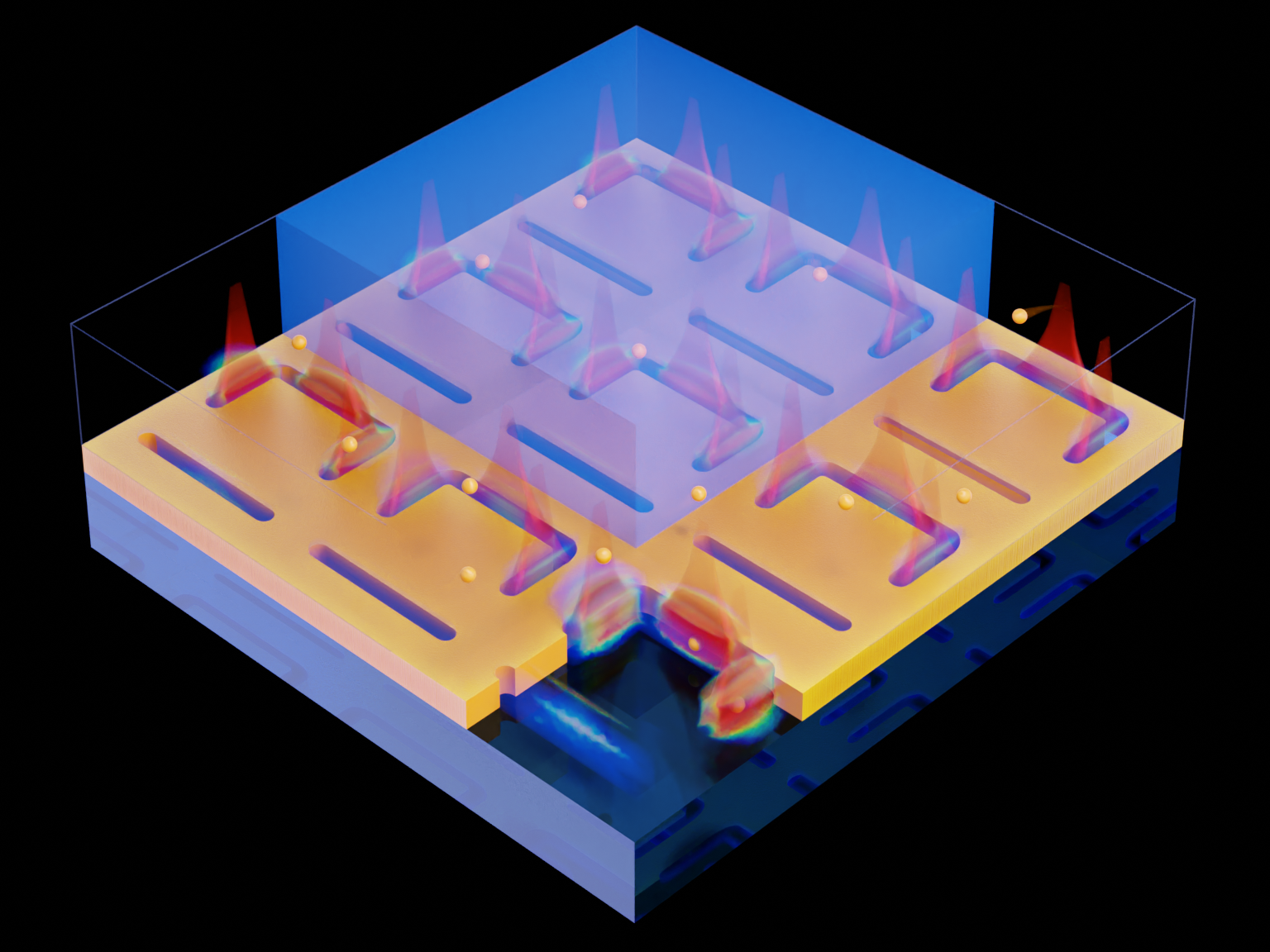}

\end{tocentry}
\singlespacing
\begin{abstract}
   The pursuit for efficient nanoparticle trapping with low powers has led to optical tweezers technology moving from the conventional free-space configuration to advanced plasmonic tweezers systems. However, trapping nanoparticles smaller than 10~nm still remains a challenge even for plasmonic tweezers. Proper nanocavity design and excitation has given rise to the self-induced back-action (SIBA) effect offering enhanced trapping stiffness with decreased laser power. In this work, we investigate the SIBA effect in metamaterial tweezers and its synergy with the exhibited Fano resonance. We demonstrate stable trapping of 20~nm gold particles for on-resonant and off-resonant conditions with experimental trap stiffnesses as high as 4.18~$\pm$~0.2~fN/(nm$\cdot$~mW/$\mu$m$^{2}$) and very low excitation intensity of about 1~mW/$\mu$m$^2$. Simulations  reveal the existence of two different groups of hotspots per unit cell of the metamaterial array. The two hotspots exhibit tunable trap stiffnesses, a unique feature that can allow for sorting of particles and biological molecules based on their characteristics.
 
\end{abstract}

\noindent{\bf Keywords:} Metamaterial tweezers, self-induced back-action, plasmonic tweezers, gold nano-particle trapping, Fano resonance

\section{Introduction}
Since the invention of optical tweezers by Arthur Ashkin\cite{RN2, Ashkin1986} numerous trapping platforms have been extensively studied and implemented for a variety of applications\cite{Polimeno, Bouloumis2020FromFT} with trapping specimens ranging from metallic particles\cite{Crozier2009, Zhang2010, Cichos2022} through biological samples\cite{Righini2009, Verschueren2019, Peng2021}, to quantum dots\cite{Xu2019AlldielectricNF, Hong2021ElectrothermoplasmonicTA, Jiang2021SinglePS}. One major direction in optical trapping is to develop a platform that can efficiently trap nanoparticles that are tens of nanometers or smaller in size, using as low a light intensity as possible, thereby minimizing heating of the specimen\cite{Roxworthy2012ApplicationOP, Jiang2020QuantifyingTR, Kotsifaki2022TheRO}. The most common configurations that meet these requirements are based on metallic nanostructures in which one can make use of the strong near-field forces that arise due to the excitation of surface plasmons~\cite{Juan2011PlasmonNT, Saleh2012TowardEO, Kotsifaki2019PlasmonicOT, Bouloumis2020FastAE}. 

However, the design of the nanostructure alone cannot improve the trapping stiffness beyond a certain point. This realization led to the implementation of the self-induced back-action (SIBA) trapping\cite{Juan2009SelfB}. In SIBA trapping, the plasmonic cavity resonance is tuned in such a way that the particle itself, while trapped in the cavity, plays a dynamic role in the reconfiguration of the intracavity intensity based on the spatial overlap of the particle and the cavity\cite{Neumeier2015SelfinducedBO}. This optomechanical coupling facilitates trapping of the particle with a lower light intensity, thereby relaxing the intensity requirements while providing a stable trap with less heating~\cite{Descharmes2013ObservationOB,Mestres2016Unravelling, Zhang2019OptimizationOM, Zhu2018}.

Trapping gold nanoparticles (AuNPs) is of particular interest, due to the versatility in their synthesis methods~\cite{GoldNPsSynthesisReview, Grammatikopoulos2016NPs} and their contribution in various fields~\cite{HalasNPsreview}. In 1908, rigorous theoretical work was done by Gustav Mie to mathematically describe the absorption of light by a subwavelength gold sphere, based on Maxwell's equations~\cite{Mie1908}. Since then, owing to the great tunability of plasmon resonances and their unique properties, AuNPs have contributed to the development of fields within physics and material science such as the surface-enhanced Raman scattering (SERS)~\cite{SERS1978} technique, enhanced photovoltaic conversion efficiency via embedded particles in active layers of solar cells~\cite{NOTARIANNI201423}, and quantum nanophotonic systems coupled to single photon emitters resulting in ultrabright sources of nonclassical light~\cite{Nguyen2018NanoassemblyOQ}. There are numerous applications in other fields, too, such as in chemistry for catalysis purposes~\cite{Lichen2018} and gas sensing~\cite{Wuenschell2022GoldSensing}, in computational methods with the newly established field of neuromorphic computing via AuNPs cluster assemblies~\cite{Mirigliano2021Neuromorph}, and in biomedical research with applications in drug and gene delivery~\cite{Ghosh2008GoldNI}, cancer treatment with photothermal therapy~\cite{Vines2019GoldNF}, vaccine development~\cite{CarabineiroGoldVaccines}, and many others~\cite{Daraee2016ApplicationOG, Nejati2021BiomedicalAO}. All these promising applications have triggered interest in trapping AuNPs of various sizes and investigations into their optomechanical properties during manipulation and nanopositioning~\cite{Crozier2009, Min2013PlasmonicTrapping, Mestres2016Unravelling}.

In our previous work~\cite{Kotsifaki2020FanoResonantAM}, we demonstrated single trapping of 20~nm diameter polystyrene particles using an array of metamolecules. The resulting trapping stiffness was measured to be 8.65~fN/(nm$\cdot$mW) and is the highest reported normalised trapping stiffness~\cite{Kotsifaki2020FanoResonantAM}. Implementing metamaterial structures in a plasmonic trapping configuration offers the advantage of being able to excite and tune a very sharp Fano resonance, which is a result of destructive interference between a broadband dipole resonance and a sharp quadrupole resonance~\cite{Fedotov2007}. This destructive interference leads to a nanocavity with very small mode volume, offering very strong confinement of particles in the trap~\cite{Fedotov2007,Papasimakis2009,Tanaka2010}. A Fano resonator is also very sensitive to changes of the local refractive index~\cite{Boris}, making this platform an ideal candidate for combining with  SIBA trapping. Proper structure design could foster a synergy between the Fano-resonance and the SIBA effect, resulting in fast optomechanical responses that could facilitate long duration trapping events with low laser power.

In this work, we experimentally trapped AuNPs with a diameter of 20~nm using the metamaterial tweezers introduced earlier~\cite{Kotsifaki2020FanoResonantAM}. Theoretical analysis of the excited plasmonic hotspots revealed a wavelength-dependent trapping stiffness per hotspot. This offers the ability to tune the hotspots' strength, with applications in sorting nanoparticles of different sizes and refractive indices. Additionally, we investigated the synergy between the metamaterial tweezers and the SIBA effect and experimentally observed SIBA-assisted trapping. To the best of our knowledge, there is no prior reporting of this in the literature. A very high trapping stiffness was obtained for on-resonant excitation (i.e., SIBA-assisted trapping), as well as for blue-detuned excitation, with the highest trapping value being  4.18~$\pm$~0.2~fN/(nm$\cdot$mW/$\mu$m$^{2})$. This value is comparable to trapping of 20~nm polystyrene particles with metamaterial tweezers where the SIBA effect could not be utilized~\cite{Kotsifaki2020FanoResonantAM}.

\section{Results and discussion}

A plasmonic metamaterial nanostructured device was used for the trapping experiments. The device was fabricated using focused ion beam milling (FIB) on a gold film, with the geometrical characteristics determined in advance through finite element method simulations. AuNPs with a diameter of 20~nm were dispersed in a heavy water/buffer mixture solution and a small amount of surfactant was also used. The solution was placed onto the device, and the sample was mounted on a custom-made microscope setup. The excitation laser was incident from the glass slide of the sample. Detailed description of the methods can be found in S1. 

\begin{figure}
     \centering \captionsetup{justification=raggedright,singlelinecheck=false}
     \begin{subfigure}[t]{0.48\textwidth}  
     \caption{}
         \centering
         \includegraphics[width=\textwidth]{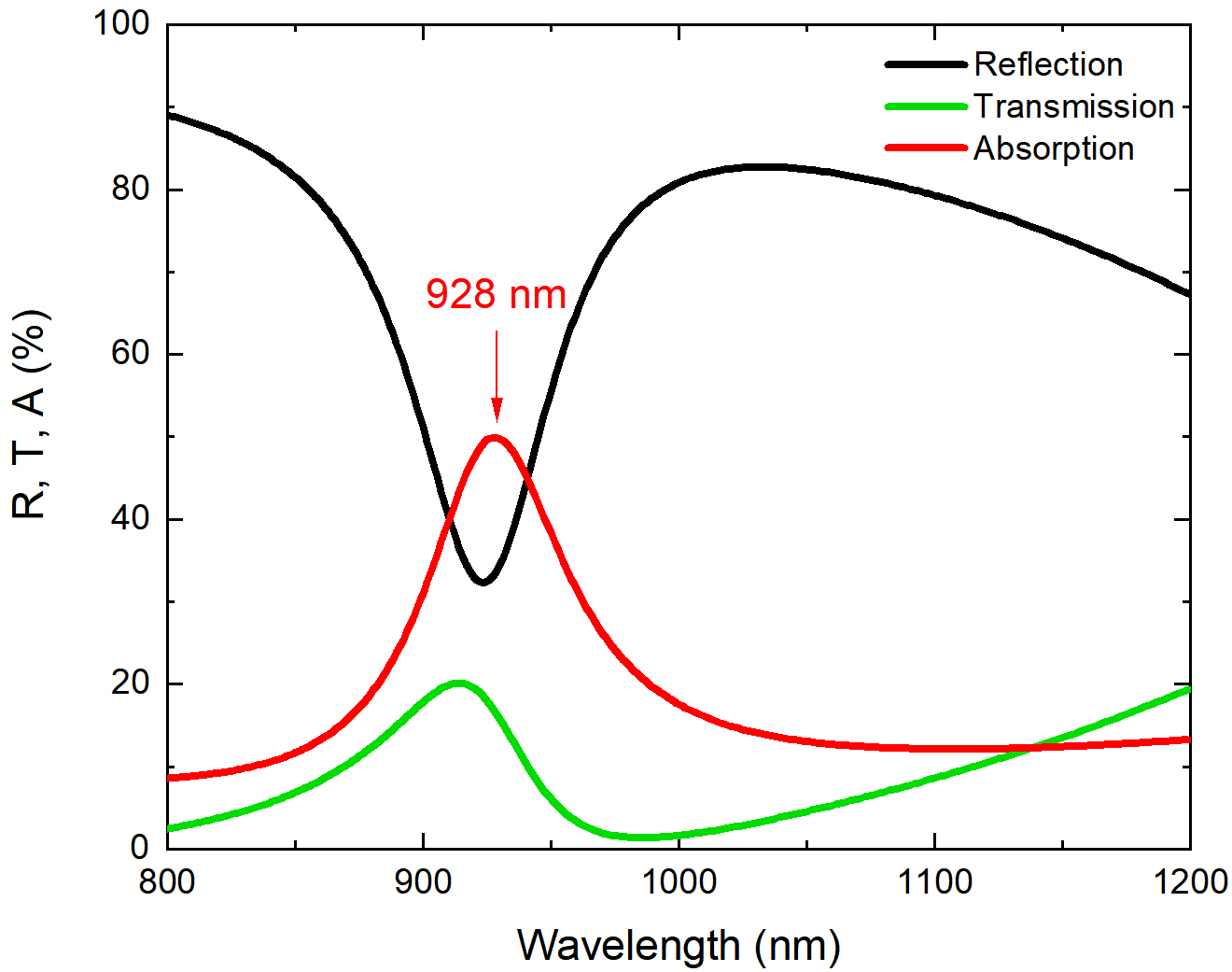}
         \label{Fig1a}
     \end{subfigure}
     \hfill
     \begin{subfigure}[t]{0.45\textwidth}
         \caption{}
         \centering
         \includegraphics[width=\textwidth]{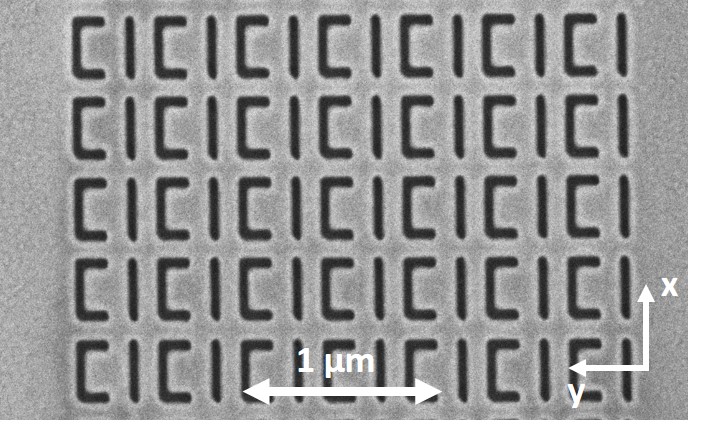}
         \label{Fig1b}
     \end{subfigure}
     \begin{subfigure}[b]{1\textwidth}
        \caption{}
         \centering
         \includegraphics[width=\textwidth]{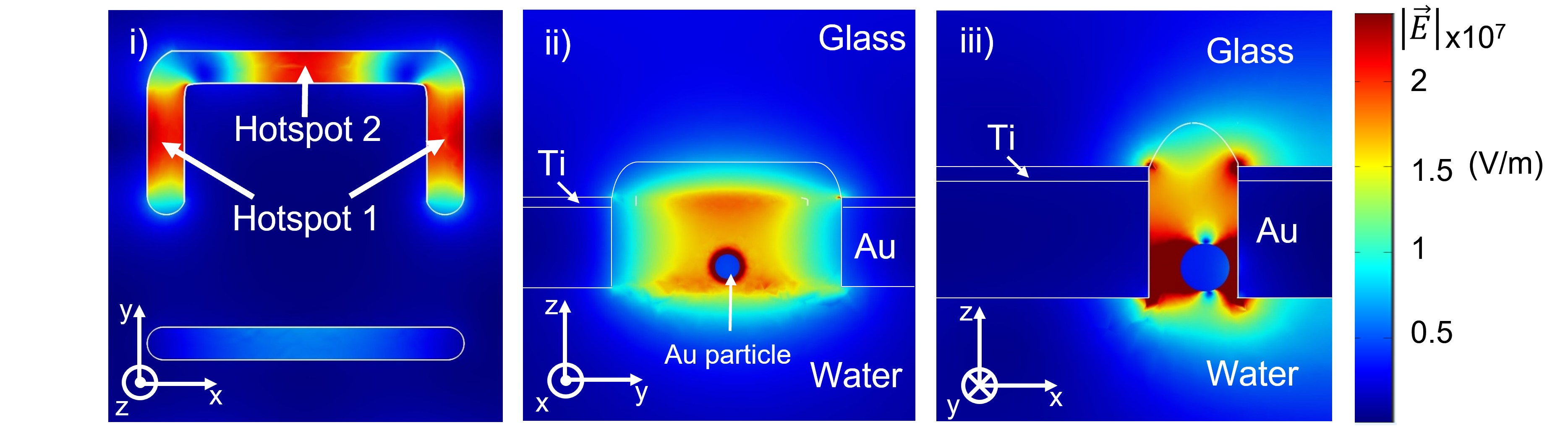}
         \label{Fig1c}
     \end{subfigure}
     \captionsetup{justification=justified}
        \caption{(a) The simulated reflection (R), transmission (T), and absorption (A) spectra of the metamaterial structure indicating the empty cavity resonance at 928~nm. (b) Scanning electron microscopy (SEM) image of the fabricated metamaterial array. (c) The electric field generated at a unit cell of the metamaterial structure, for the case of (i) an empty cavity and (ii, iii) when a particle is trapped. The fields  are for 928~nm excitation light at the planes (i) $z~=~13$~nm (ii) $x~=~149$~nm, and (iii) $y~=~75$~nm. The particle is at the equilibrium position of Hotspot 1.}
        \label{Fig1}
\end{figure}

Figure~\ref{Fig1}a shows the simulated reflection, transmission and absorption spectra of the fabricated metamaterial device (Figures~\ref{Fig1}b and \ref{SI_metanunit}). The absorption peak at 928~nm indicates the resonance of the device, hence the incident wavelength required to excite the plasmonic field, when the device is immersed in water solution. Microspectrophotometry measurements revealed that the experimental resonance peak was in perfect agreement with the simulations - since we cannot directly measure the absorption resonance experimentally, we used a formula to calculate the experimental absorption resonance peak from the transmission spectrum (see Figure~\ref{SI_structure}b). Figure~\ref{Fig1}c(i) shows the simulated empty-cavity electric field for 928~nm excitation light, propagating in the $z$-axis, and linearly polarized along the $y$-axis. Each unit cell has three hotspots where particles can be trapped; two symmetric and identical ones along the $y$-axis, from now on referred to as Hotspot 1 (we make no distinction between them) and one hotspot along the $x$-axis, from now on referred to as Hotspot 2. We assume that a particle moving under Brownian motion in the solution has equal probabilities of being in the vicinity of any of the two hotspots and subsequently trapped. In Figures~\ref{Fig1c}(ii),(iii) the electric field is plotted at the $yz$ and $xz$ planes while a AuNP 20~nm in diameter is trapped at the equilibrium position of the Hotspot 1 (see S4 for details on the equilibrium position). Due to its metallic nature, the particle also has a plasmonic resonance and behaves like a trapped nanoantenna, leading to significant field enhancement within the nanocavity (Figure~\ref{Fig1}c(iii)).  

\begin{figure}
    \centering
    \includegraphics[width=0.65\textwidth]{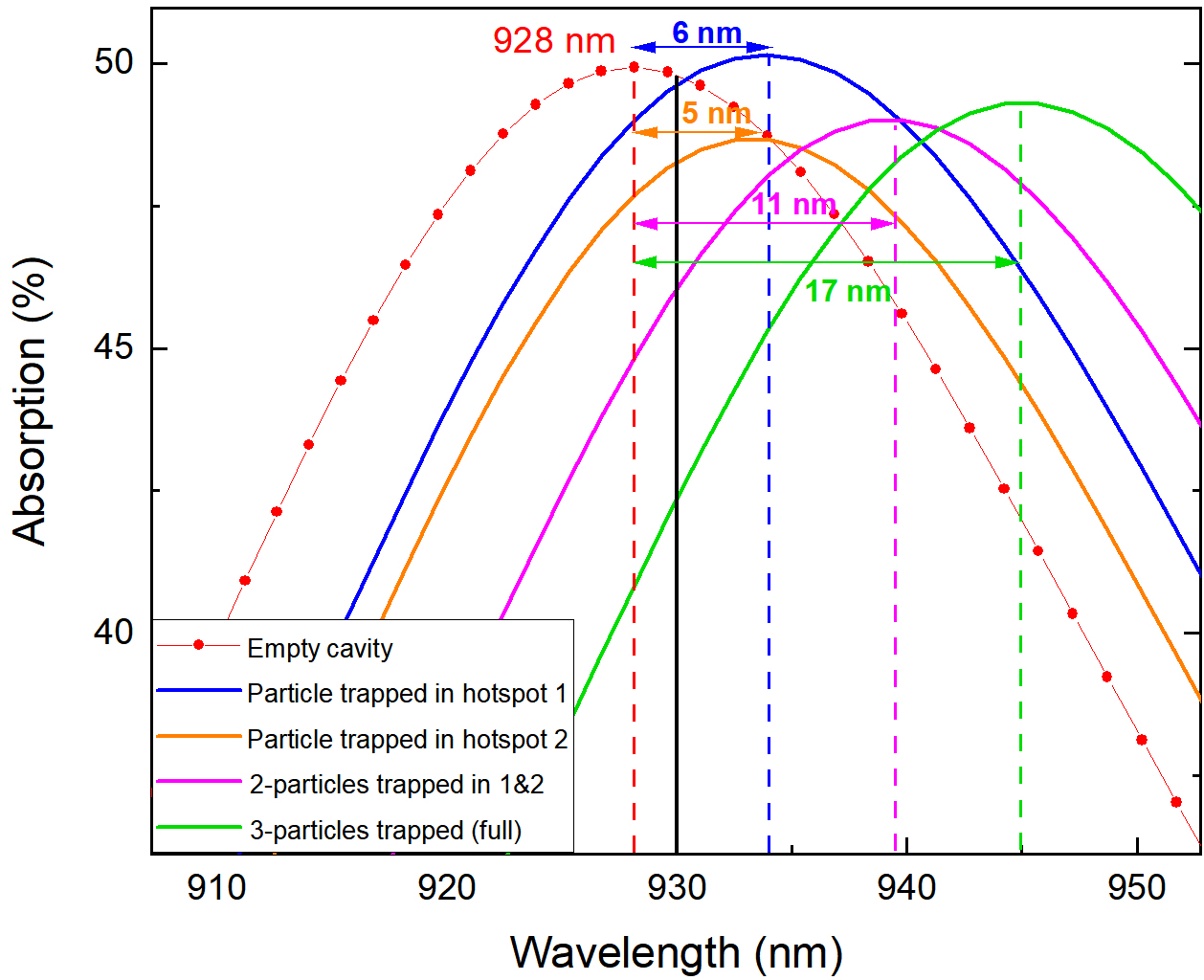}
    \caption{Simulated absorption spectra for different configurations of trapped particles. Empty cavity absorption resonance: red line with dots. Cavity shifts created by one (blue and orange lines), two (magenta line) and three (green line) particles trapped at the same unit cell in various combinations of the three hotspots. The cavity shifts from the empty cavity resonance (928~nm) are indicated by the arrows in corresponding colors.The solid black line at around 930~nm indicates the wavelength around which there is a symmetrical cavity shift between the empty cavity resonance and the shifted cavity resonance resulting from a single particle trapped.}
    \label{Fig2}
\end{figure}

The existence of a trapped particle inside the nanocavity not only creates an enhancement of the \textit{E}-field, but also results in a frequency shift of the cavity resonance given by~\cite{Mestres2016Unravelling} 
\begin{equation}
    \delta \omega (r_p)=\omega_c  \frac{\alpha_d}{2V_m\epsilon_0}f(r_p)
\end{equation}

\noindent where $\omega_c$ and $V_m$ are the resonant frequency and the mode volume of the empty cavity, respectively, $\alpha_d$ is the polarizability of the nanoparticle, $\epsilon_0$ is the vacuum permittivity, and $f(r_p)$ is the cavity intensity profile. Using Equation 1, we calculated the expected cavity shift in our configuration to be $\sim$4.06~nm (see S2). This is of the same order of magnitude obtained through simulations, as shown in Figure~\ref{Fig2}. For a particle trapped in either Hotspot 1 or Hotspot 2 the resulting cavity redshift was 6~nm and 5~nm, respectively, and for consecutive trapping events the cavity shifts by an additional step of 6~nm or 5~nm depending on where the particle was trapped. This small discrepancy between the analytical and numerical shift value is attributed to the approximate formula that was used to calculate the mode volume~\cite{Tanaka2010}, $V_m$. Based on the simulated value of the cavity shift, we estimated that our system provides an optomechanical coupling constant of approximately $G=2\pi\cdot71$~GHz/nm (see S2), which is of the same order of magnitude as other works~\cite{Thijssen2015Plasmomechanical, Mestres2016Unravelling}. The optomechanical constant defines the response of the cavity (intracavity photon flux) due to the motion of the particle in the trap and subsequent changes of momentum. It is thus important for SIBA trapping to have as high an optomechanical coupling constant as possible to achieve strong confinement of the particle. It is expected that smaller particles would have a lower optomechanical coupling constant due to a smaller change of momentum; however, in our case, the optomechanical coupling value was very high considering the small size of the particle. We attribute this to the Fano resonator being very sensitive to the motion of the particle (changes of the local refractive index).

The particle-enabled cavity shift is the principle upon which the SIBA effect is based. In the simplest case of one trapped particle, to enable the SIBA effect one should excite the metamaterial with light at 930~nm. This is approximately the wavelength around which the cavity shift is symmetrical to the empty cavity resonance (Fig.~\ref{Fig2} black vertical line). Under this condition, the particle would be trapped efficiently with a lower laser intensity and with its Brownian motion restricted by the intracavity field increasing or reducing in a dynamical way according to its motion~\cite{Juan2009SelfB}.The additional fast optomechanical response of the Fano nanocavity makes this synergy between the metamaterial tweezers and SIBA ideal for trapping nanoparticles.


\begin{figure}
     \centering \captionsetup{justification=raggedright,singlelinecheck=false}
     \begin{subfigure}[c]{0.6\textwidth}
         \centering
         \caption{}
         \includegraphics[width=\textwidth]{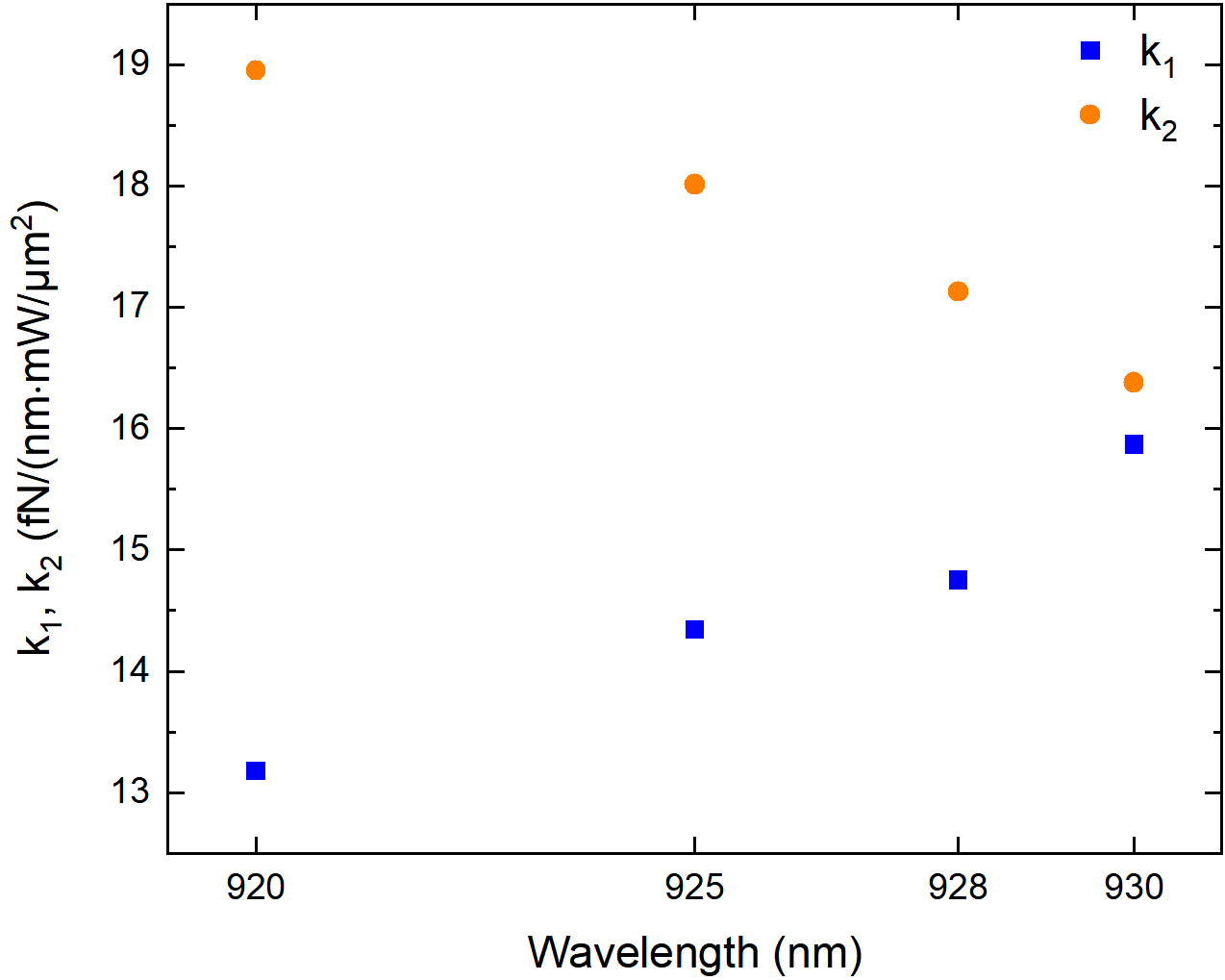}
         \label{Fig3a}
     \end{subfigure}
    \hfill 
     \begin{subfigure}[c]{0.6\textwidth}
         \centering
         \caption{}
         \includegraphics[width=\textwidth]{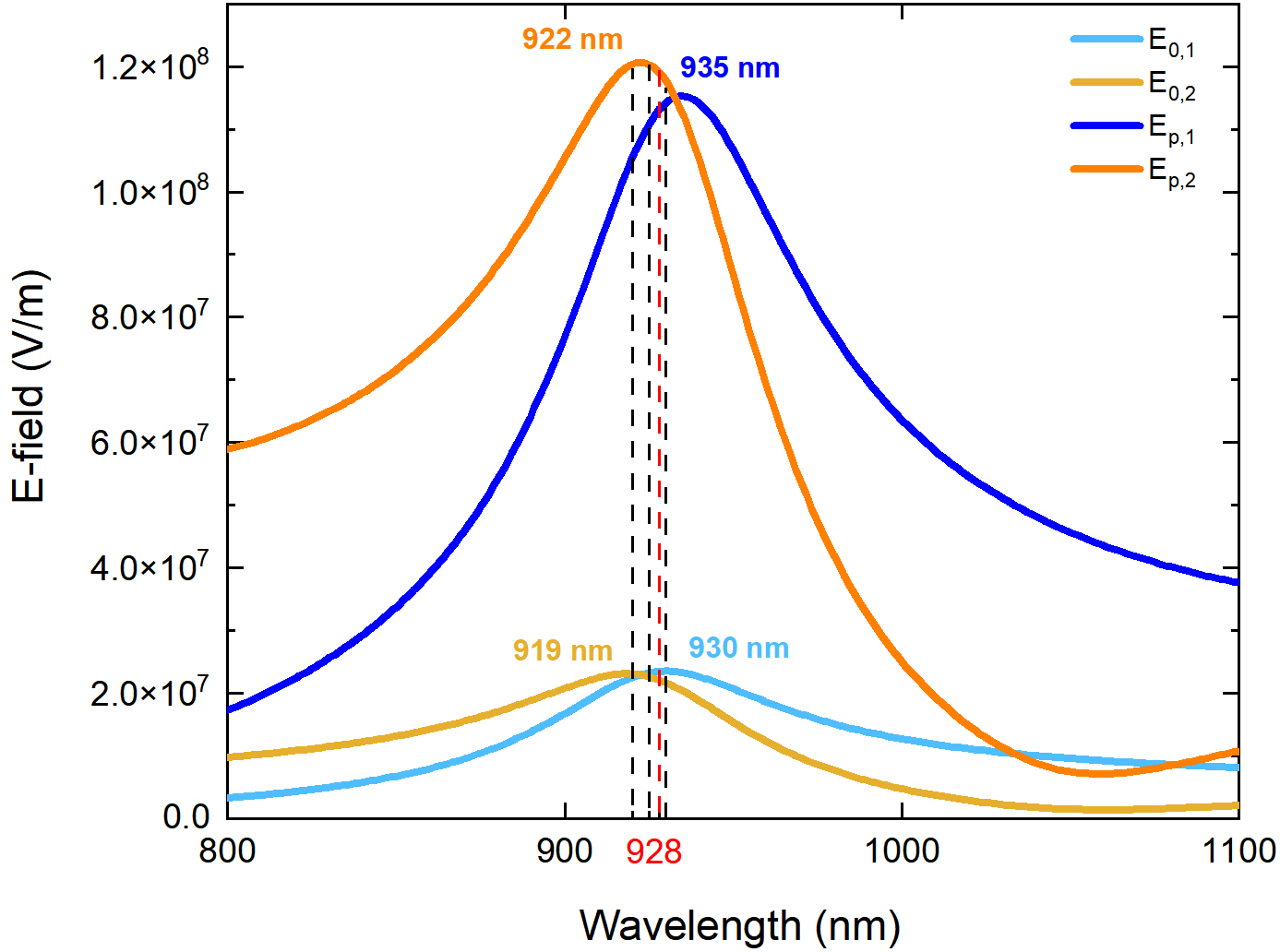}
         \label{Fig3b}
     \end{subfigure}
     \captionsetup{justification=justified}
        \caption{(a) Simulated trapping stiffnesses per wavelength for particles trapped in Hotspot 1 (k$_{1}$ - blue points) or Hotspot 2 (k$_{2}$ - orange points) calculated from the associated optical forces (see S4). (b) Electric field as a function of the excitation wavelength for the empty cavity case at the center of Hotspot 1 (E$_{0,1}$) and Hotspot 2 (E$_{0,2}$), and when a particle is trapped E$_{p,1}$ and E$_{p,2}$, respectively. 
        }
        \label{Fig3}
\end{figure}

To further map the optical forces exerted on the nanoparticles around the hotspots and identify the equilibrium positions of the trapped particles, extensive simulations were performed. In Figure~\ref{Fig3a} the trapping stiffnesses for the two hotspots are plotted as derived from the simulated optical forces (see S4 for the simulated optical forces and potentials). An interesting behavior is observed, where the trapping stiffnesses around Hotspot 1 are increasing as we scan the excitation wavelength from 920-930~nm, but an opposite, decreasing trend is observed for the stiffnesses around Hotspot 2 (same trends observed for the optical forces and potential depths, respectively).
Additionally, particles trapped at Hotspot 2 experience higher trapping stiffness, $k_2$, than particles trapped at Hotspot 1, $k_1$, for every wavelength. However, the two hotspots exhibit similar trapping stiffnesses when excited with 930~nm light. 

To interpret this behavior, we simulated the \textit{E}-field magnitudes at the two hotspots for both an empty cavity and a trapped particle (Fig.~\ref{Fig3b}). Despite the simulated absorption resonance being at 928~nm (Fig.~\ref{Fig2}), Hotspot 1 exhibits a maximum \textit{E}-field intensity at 930~nm, while Hotspot 2 has a maximum at 919~nm when the cavity is empty. Additionally, the maximum \textit{E}-field intensity for both hotspots has a similar magnitude. However, the situation is different when a particle is trapped. The presence of the metallic particle in the plasmonic nanocavity creates an enhancement of the \textit{E}-field by approximately 6 times (Fig.~\ref{Fig3b}). The total \textit{E}-field in the cavity is stronger for Hotspot 2 than for Hotspot 1. This is because the slot width at Hotspot 2 ($w_2~\approx~33$~nm) is smaller than the slot width at Hotspot 1 ($w_1~\approx~39$~nm), see Figure~\ref{SI_metanunit}. This leads to a stronger \textit{E}-field and better confinement of the nanoparticle (Fig.~\ref{SI_hotspots}). This also explains the higher trapping stiffness at Hotspot 2.

The increasing and decreasing trend of the trapping stiffnesses at Hotspots 1 and 2 (Fig.~\ref{Fig3a}) also follows the generated \textit{E}-field trend.  It is related to the different \textit{E}-field resonance values for each hotspot, as indicated by the dashed lines in Figure~\ref{Fig3b}. This offers the ability to excite the hotspots with a different ratio of resulting trapping stiffnesses, hence the possibility of sorting particles based on their size, shape, and refractive index at different hotspots, e.g. bigger particles could be trapped in the hotspot with the lowest trapping force and larger slot width, while smaller particles that need higher optical forces could be trapped in the other hotspot. The optical forces as a function of excitation wavelength are shown in Figure~\ref{SI_F-lamda} and exhibit the same resonance features as the \textit{E}-field.


\begin{figure}
     \centering \captionsetup{justification=raggedright,singlelinecheck=false}
     \begin{subfigure}[b]{0.4\textwidth}
         \centering
         \caption{}
         \includegraphics[width=\textwidth]{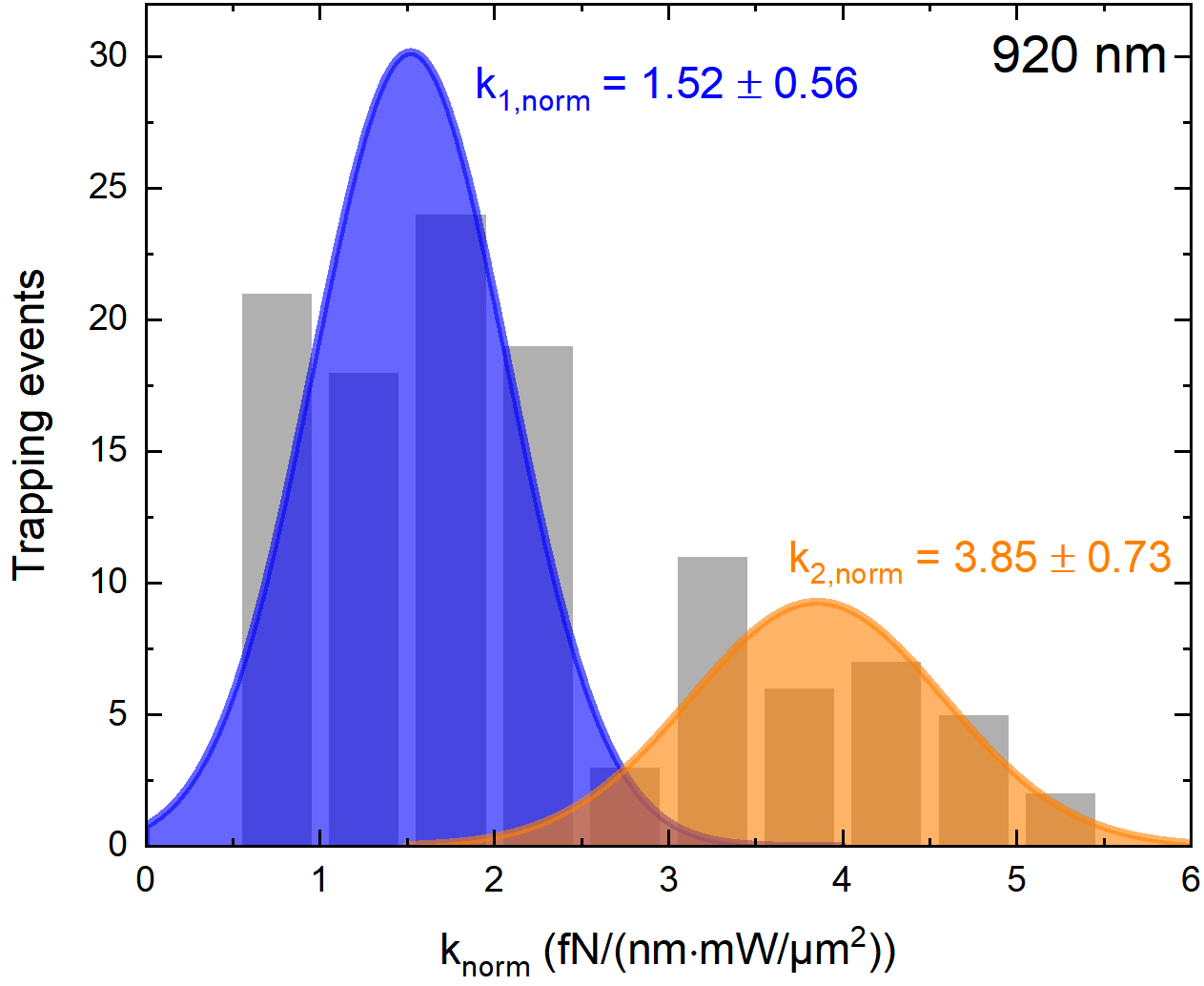}
         \label{Fig4-920}
     \end{subfigure}
     \hspace{1cm}
     \begin{subfigure}[b]{0.4\textwidth}
         \centering
         \caption{}
         \includegraphics[width=\textwidth]{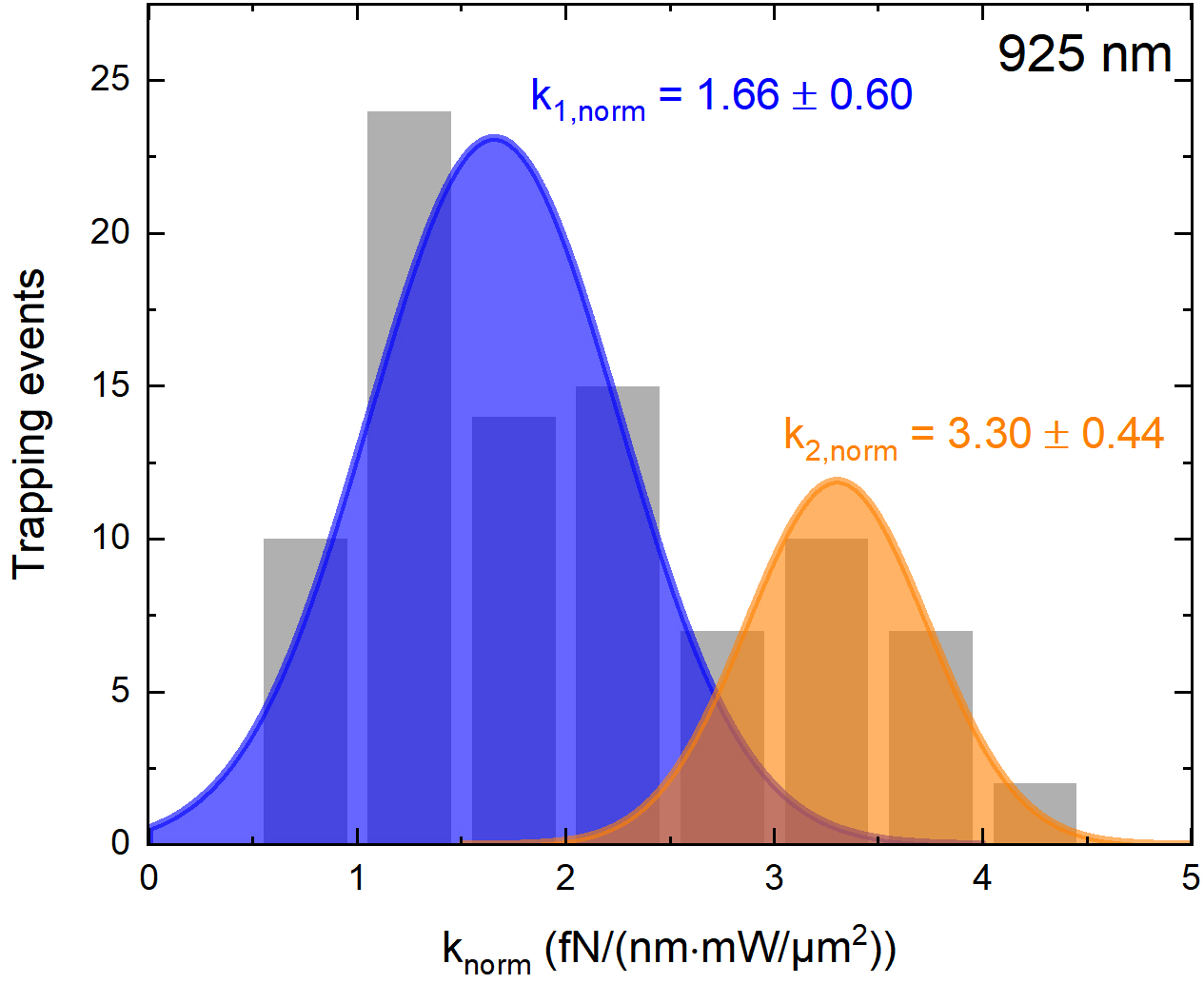}
         \label{Fig4-925}
     \end{subfigure}
     \hfill
     \begin{subfigure}[b]{0.4\textwidth}
         \centering
         \caption{}
         \includegraphics[width=\textwidth]{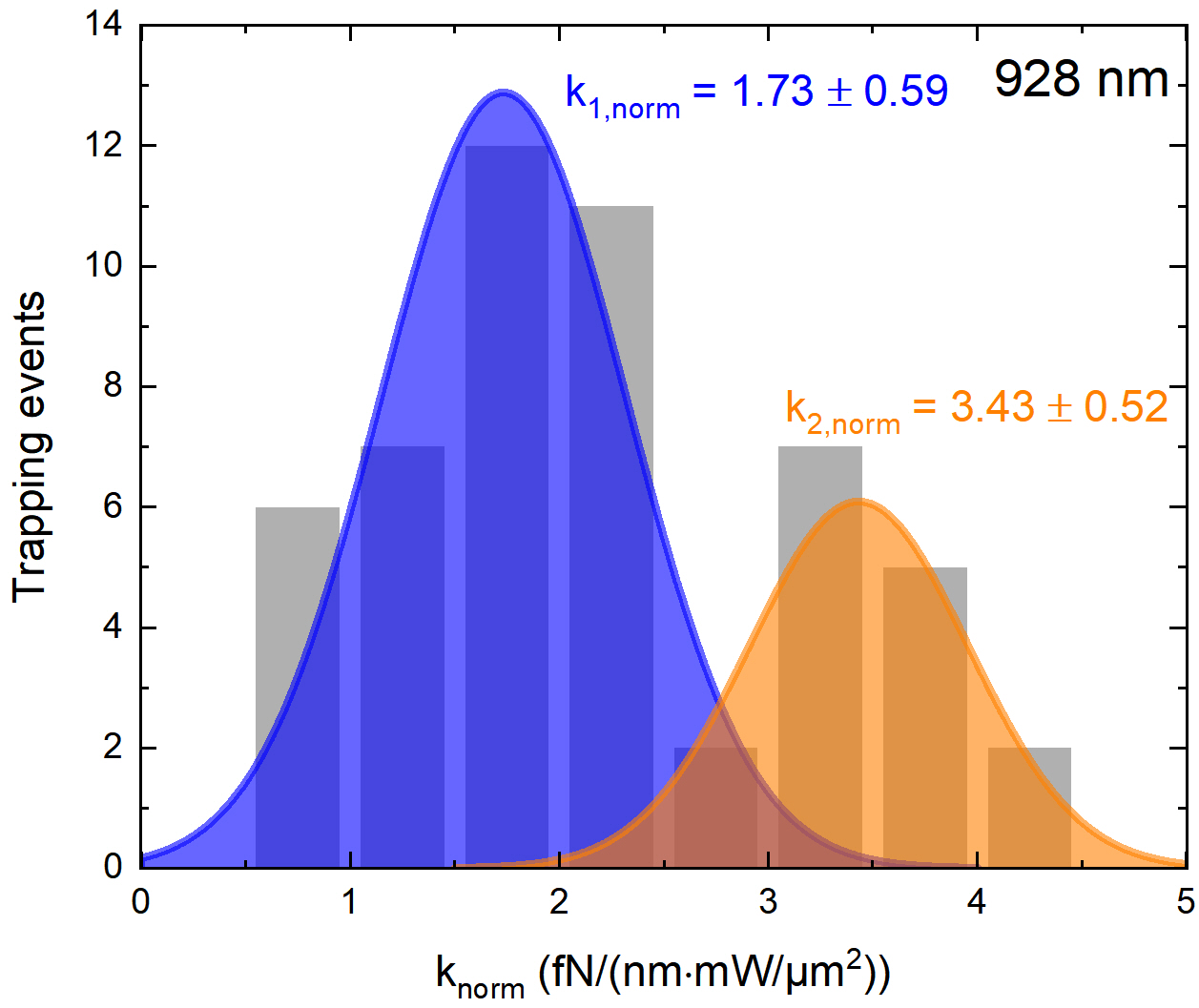}
         \label{Fig4-928}
     \end{subfigure}
     \hspace{1cm}
     \begin{subfigure}[b]{0.4\textwidth}
         \centering
         \caption{}
         \includegraphics[width=\textwidth]{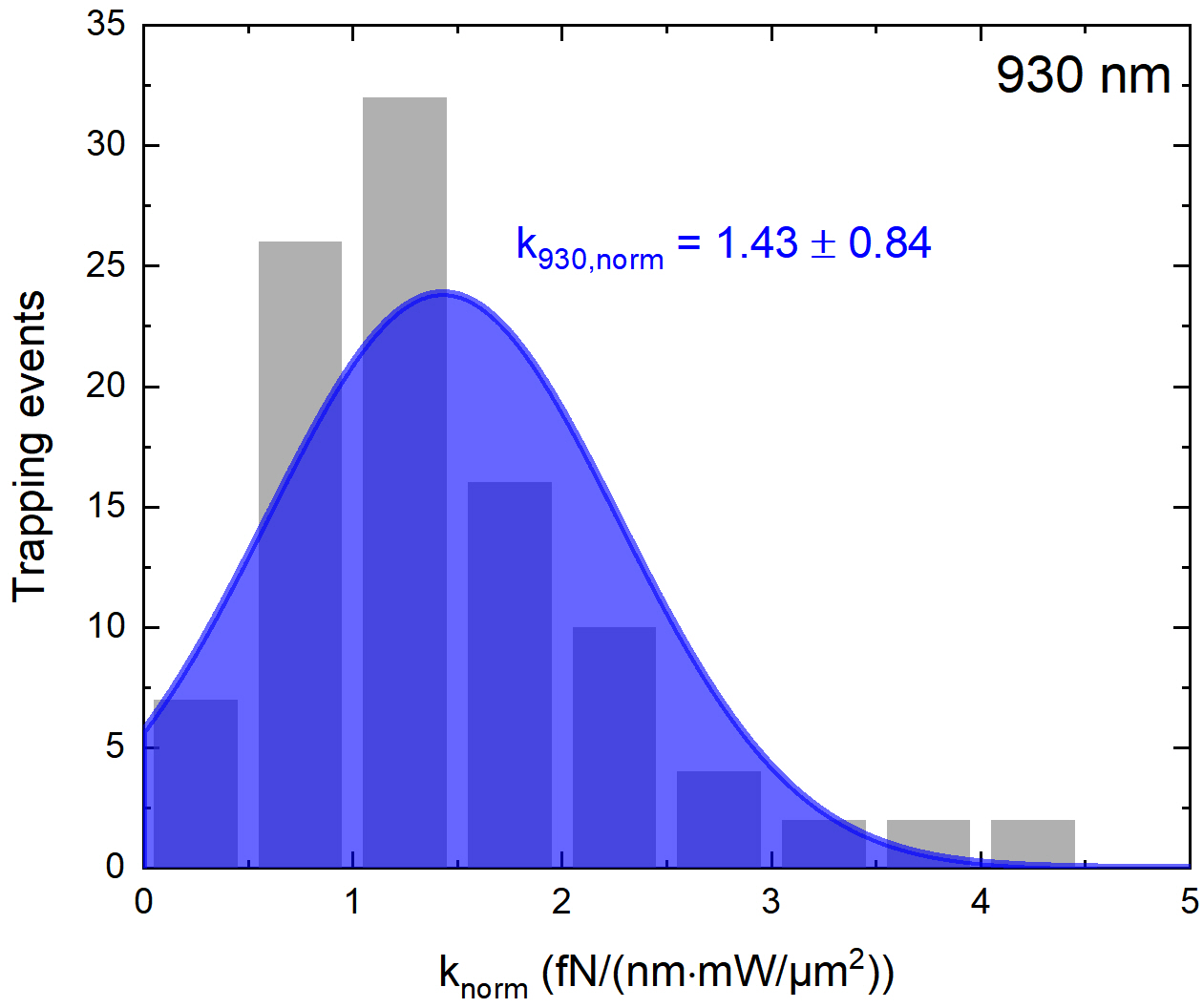}
         \label{Fig4-930}
     \end{subfigure}
    \hfill
     \begin{subfigure}[b]{0.35\textwidth}
         \centering
         \caption{}
         \includegraphics[width=\textwidth]{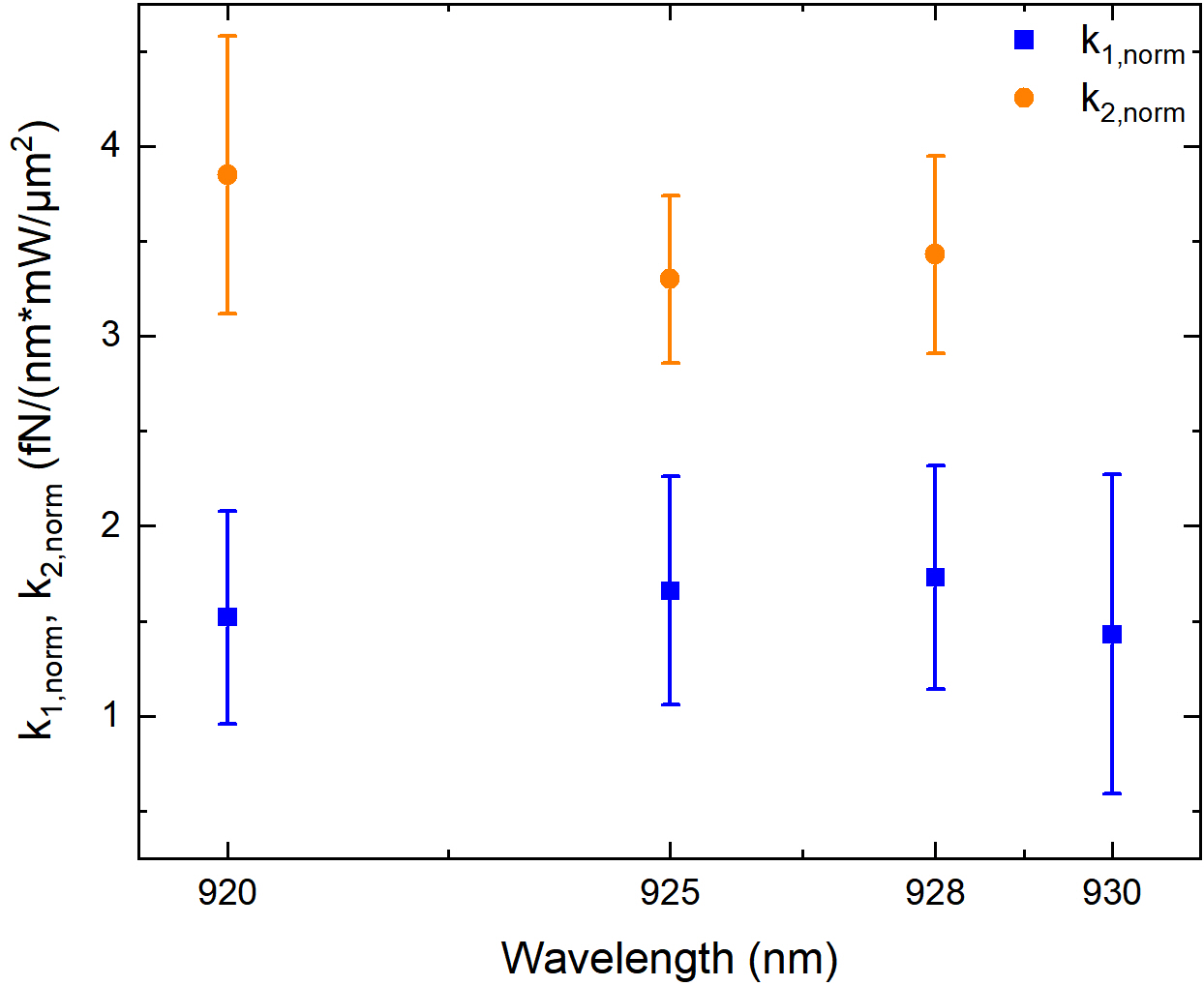}
         \label{Fig4-kexp}
     \end{subfigure}
        \captionsetup{justification=justified}
        \caption{Probability distributions of the normalized experimental trapping stiffnesses for the excitation wavelengths (a) 920~nm, (b) 925~nm,(c) 928~nm, and (d) 930~nm, with the Gaussian fits corresponding to trapping events at the Hotspot 1 (blue curves) and Hotspot 2 (orange curves). (e) The average normalized experimental trapping stiffnesses per excitation wavelength for the two hotspots. The values were obtained from the Gaussian fittings of the histograms above.
 }
        \label{Fig4}
\end{figure}

In experiments, we performed trapping with the same excitation wavelengths used for the simulations, with incident powers ranging from 1.7 - 4.3~mW. For each wavelength, up to 10 experiments were performed for increasing incident laser power with a step of 0.5~mW with a total number of about 100 trapping events per wavelength. The trapping events were recorded and analyzed based on the trapping transient analysis~\cite{SIMMONS19961813}, in order to derive the experimental trapping stiffness, $k_{exp}$. In Figure~\ref{Fig4} we present the probability distributions of the trapping stiffnesses normalized to the incident laser intensity. Data were fitted with a normal distribution function and, for the wavelengths 920~nm~-~928~nm, two distribution functions were required in order to fit the experimental results. This is in agreement with the simulations where particles trapped at Hotspot 1 (blue curves) experience a weaker trapping stiffness compared with particles trapped at Hotspot 2 (orange curves). However, for trapping with 930~nm wavelength (Fig.~\ref{Fig4-930}), a single distribution function is sufficient since, at this wavelength, the two hotspots have a similar trapping stiffness and are indistinguishable. This is confirmed by Figure~\ref{Fig3a}. It is also worth noting that the probability of trapping at Hotspot 1 is significantly higher than at Hotspot 2; this is  expected since each metamolecule is comprised of twice as many hotspots of type 1.

The averaged normalized experimental trapping stiffnesses, $k_{1,norm}$ and $k_{2,norm}$ (Fig.~\ref{Fig4-kexp}), although following the same trend as the theoretical ones (Fig.~\ref{Fig3a}), are approximately 5~-~8 times lower in magnitude. The maximum normalized trapping stiffness obtained was 4.18~fN/(nm $\cdot$ mW/ $\mu$m$^2$) for excitation at 920~nm which is of the same order of magnitude as that  previously reported for polystyrene particle trapping~\cite{Kotsifaki2020FanoResonantAM}. However, the deviation from the simulated values indicates the existence of additional destabilisation forces that were not considered in the simulations and may affect the trapping process.


\begin{figure}
     \centering \captionsetup{justification=raggedright,singlelinecheck=false}
     \begin{subfigure}[b]{0.49\textwidth}
         \centering
         \caption{}
         \includegraphics[width=\textwidth]{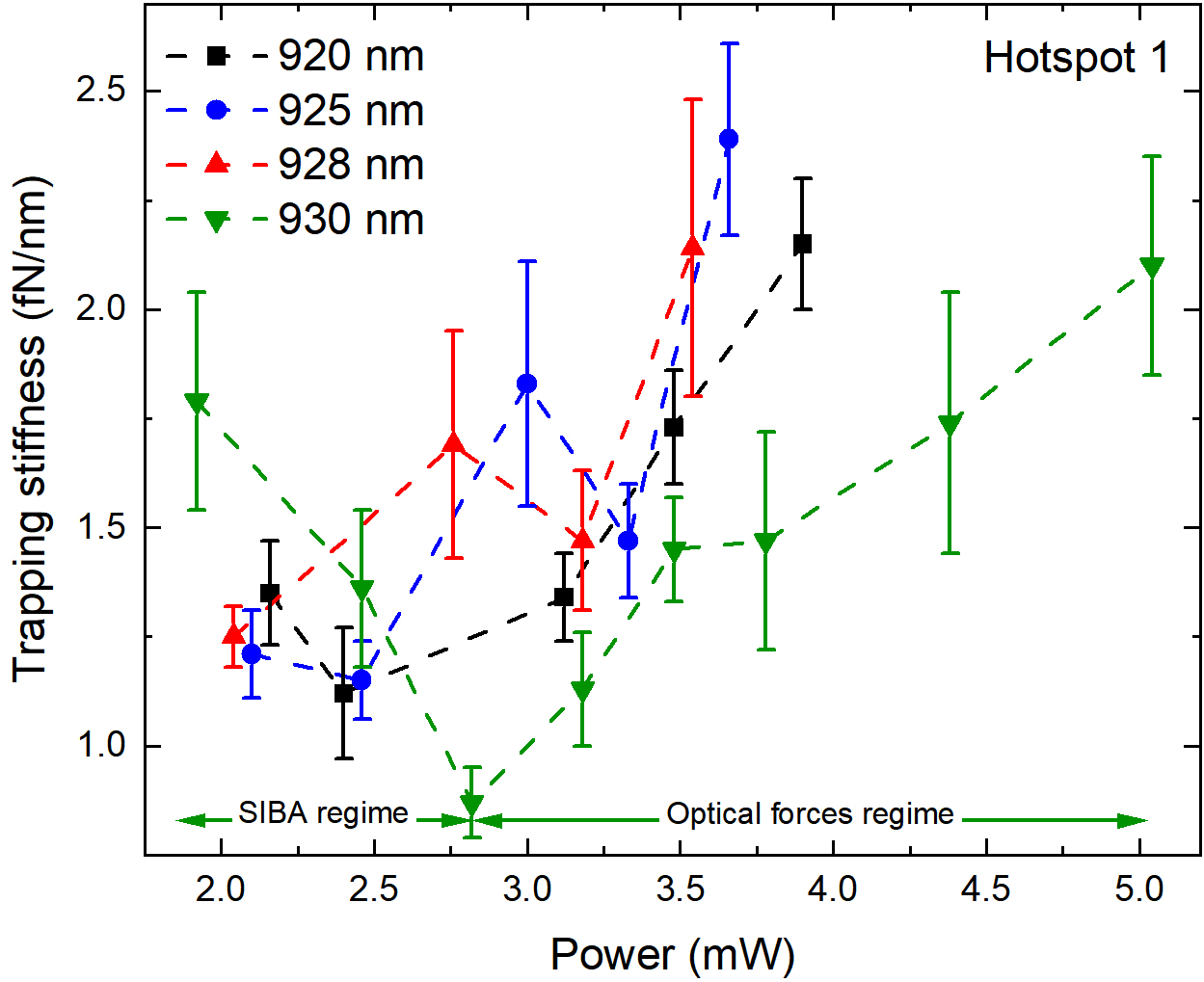}
         \label{Fig5a}
     \end{subfigure}
     \hfill
     \begin{subfigure}[b]{0.48\textwidth}
         \centering
         \caption{}
         \includegraphics[width=\textwidth]{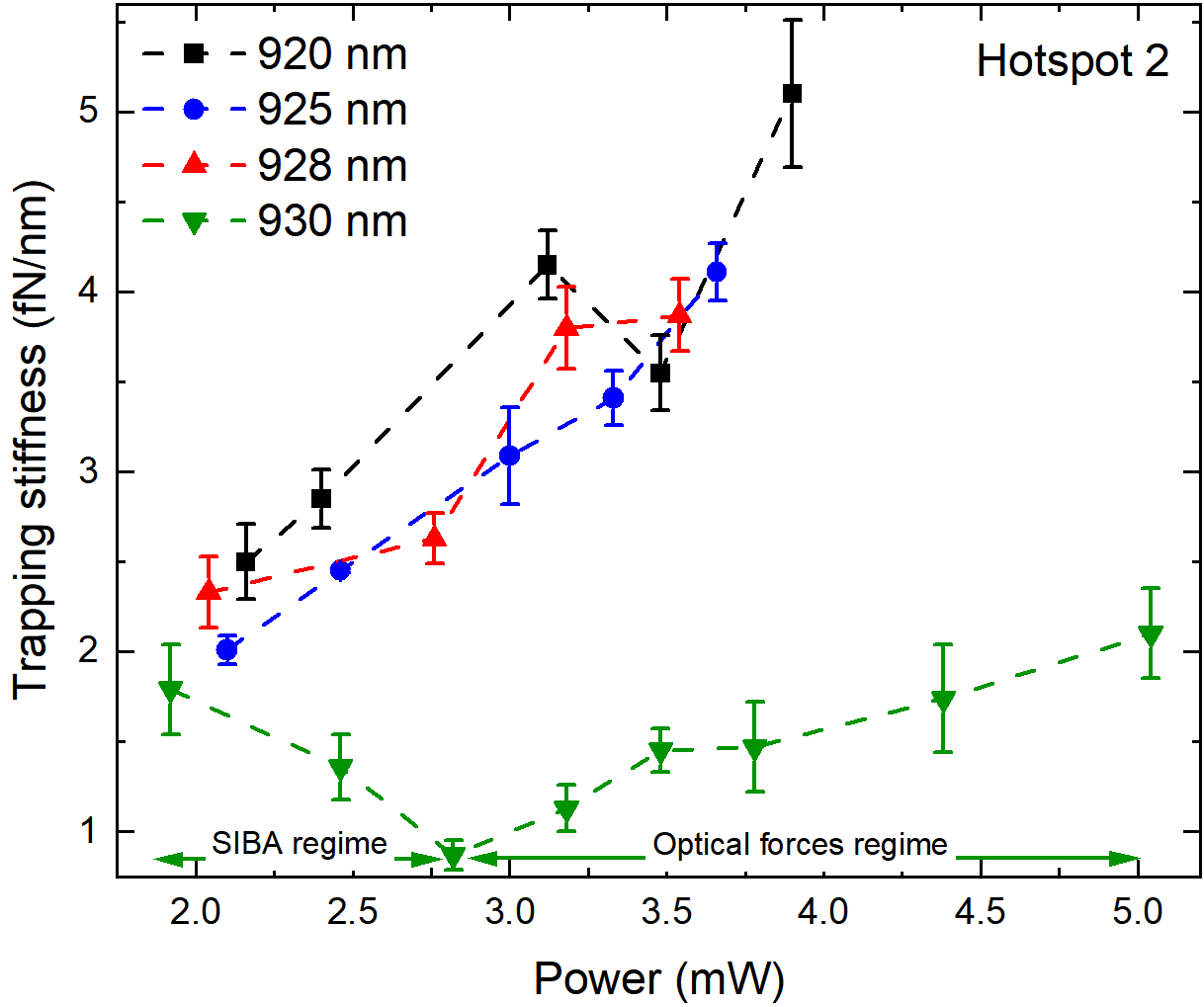}
         \label{Fig5b}
     \end{subfigure}
     \hfill
     \begin{subfigure}[b]{0.48\textwidth}
         \centering
         \caption{}
         \includegraphics[width=\textwidth]{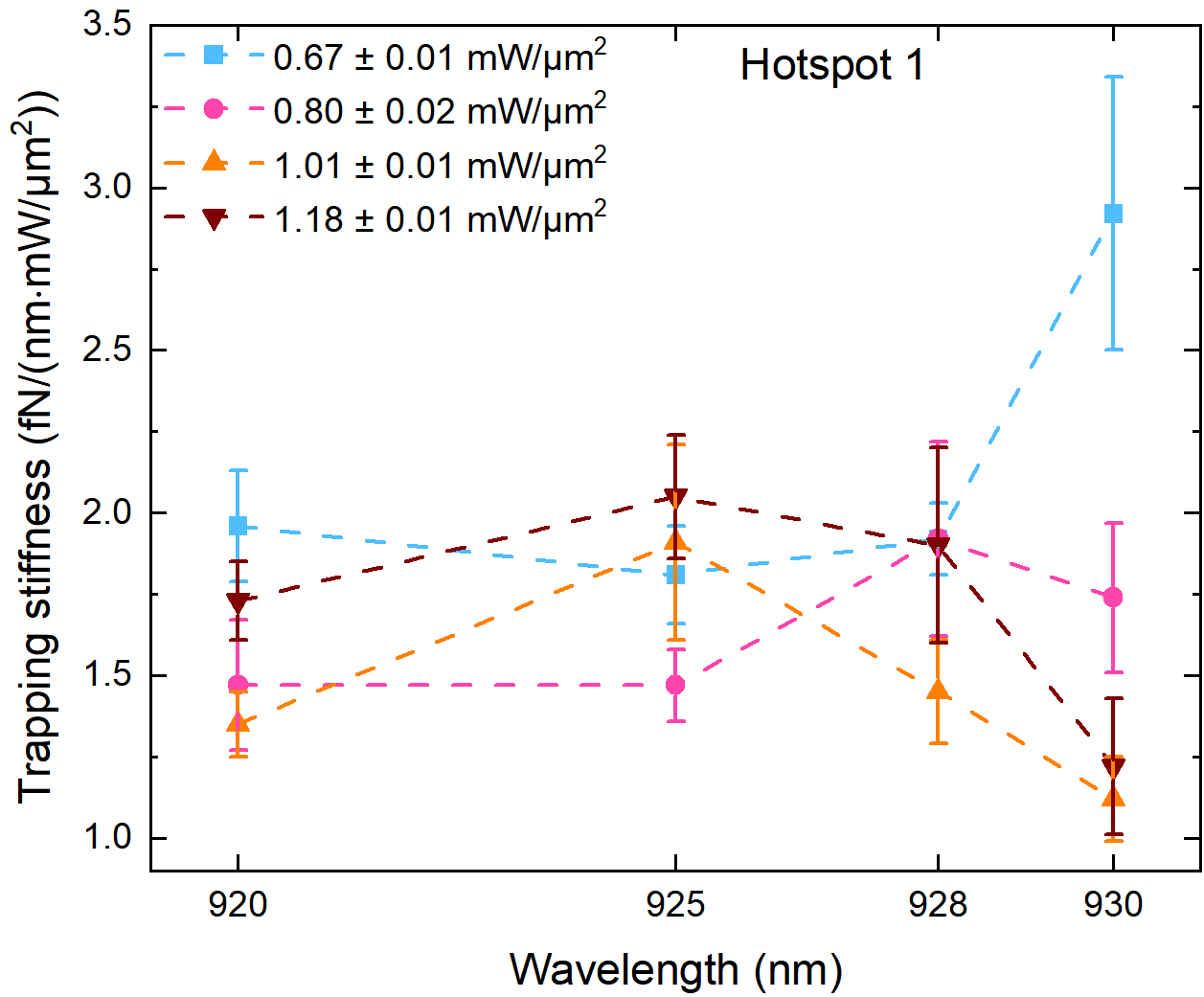}
         \label{Fig5c}
     \end{subfigure}
     \hfill
     \begin{subfigure}[b]{0.48\textwidth}
         \centering
         \caption{}
         \includegraphics[width=\textwidth]{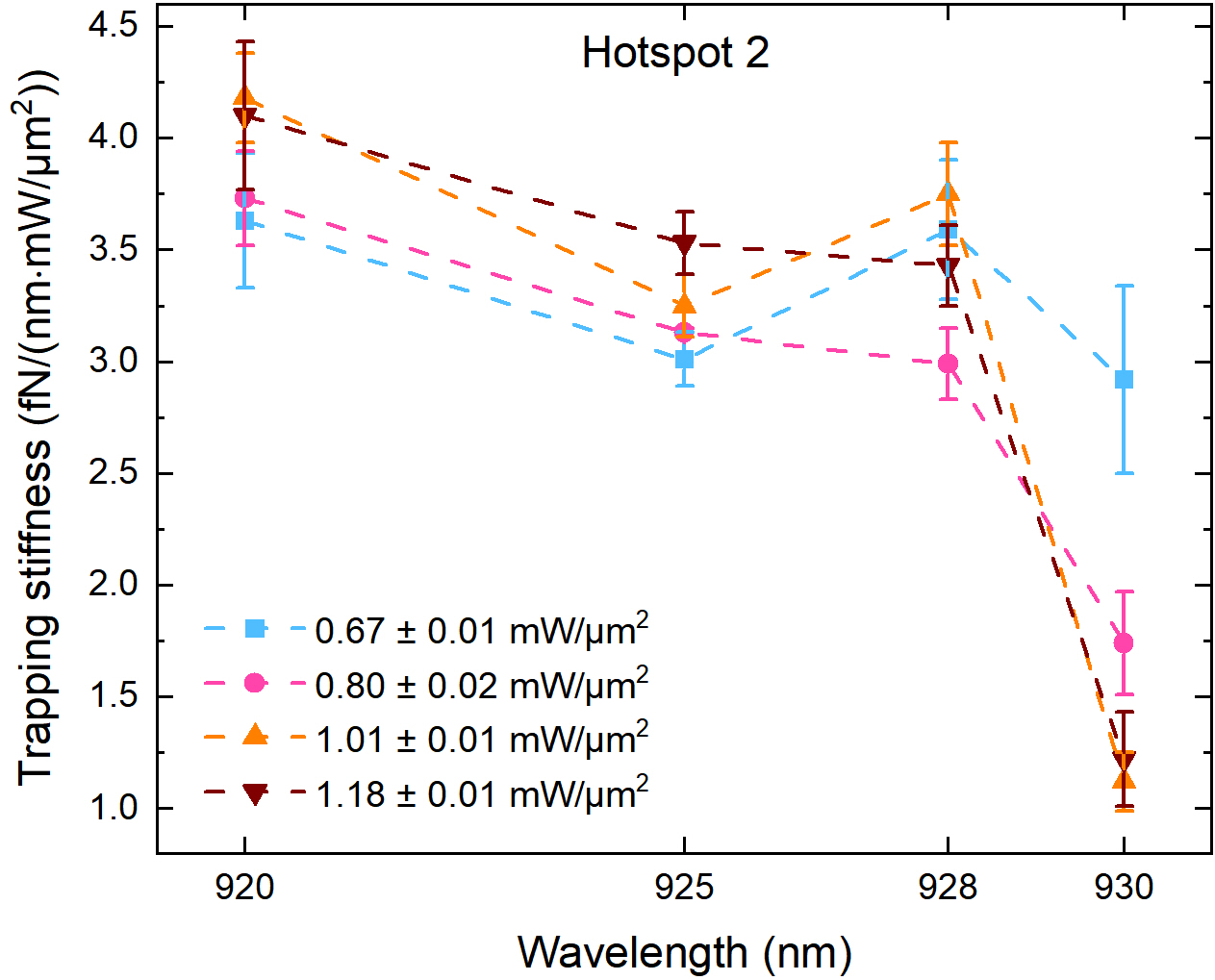}
         \label{Fig5d}
     \end{subfigure}
     \captionsetup{justification=justified}
        \caption{Experimental trapping stiffness as a function of incident laser power for particles trapped at (a) Hotspot 1  and (b) Hotspot 2, and normalized trapping stiffness as a function of the excitation wavelength, respectively (c) and (d).}
        \label{Fig5}
\end{figure}

The situation is different when trapping with the 930~nm excitation laser; the average trapping stiffness magnitude, $k_{930,norm}$, significantly deviates from the simulated behavior, exhibiting the lowest value amongst all wavelengths (Fig.~\ref{Fig4-kexp}). To explain this and get a clearer view on the experimental findings, in Figure~\ref{Fig5} we plotted the initial non-normalized trapping stiffnesses as a function of the incident laser power and trapping wavelength, based on the statistical analysis for the two hotspots. Although the behavior is quite complex, an increasing trend of the trapping stiffness with increasing incident laser power is observed for the wavelengths 920~nm~-~928~nm for both hotspots (Fig.~\ref{Fig5}a,b). This trend is clearly not linear, as theoretically predicted, and it is attributed to the collective effect of different forces acting on the nanoparticles. Interestingly, the behavior of the 930~nm wavelength is less complicated and in agreement with previous works where the SIBA effect was investigated~\cite{Mestres2016Unravelling}.

As mentioned earlier, exciting the structure with 930~nm light is strongly desired because at this wavelength the SIBA contribution is enabled, resulting in a high trapping stiffness with a low laser intensity. Indeed, as shown in Figure~\ref{Fig5}a,b, trapping with 930~nm light results in two distinct regimes of trapping stiffness values. These are the result of a combination of the optical forces along with the SIBA effect acting on the nanoparticle. The highest normalized trapping stiffness when trapping with 930~nm was measured to be 2.92~fN/(nm$\cdot$mW/$\mu$m$^{2}$) (Fig.~\ref{Fig5}c,d) for an incident intensity as low as 0.61~mW/$\mu$m$^{2}$. In the low power regime, the SIBA effect is the dominant mechanism keeping the particle trapped since the optical forces are not strong enough~\cite{Neumeier2015SelfinducedBO}. As the incident power increases, the optical forces play a more dominant role, and the trapping stiffness decreases because the scaling between laser power and particle confinement, upon where the SIBA is based, becomes less efficient~\cite{Neumeier2015SelfinducedBO}. After a certain power threshold, which in our experiment lies at 2.82~mW~(1 mW/$\mu$m$^{2}$ intensity), the SIBA effect vanishes and the optical forces linearly scale up with the power. In principle, based on Figures~\ref{Fig5}c,d, we observe that low optical intensities result in higher trapping stiffnesses when exciting on-resonant for the SIBA (930~nm); however, for higher intensities  the blue-detuned excitation of the cavity provides stiffer traps.

At Hotspot 2, due to better confinement of the \textit{E}-field, the resulting plasmonic field was stronger than that at Hotspot 1. Even for low powers and off-resonant conditions, high optical forces were exerted on the nanoparticles (Fig.~\ref{Fig5}b,d). Despite the slightly higher trapping stiffnesses that were obtained when exciting at off-resonant conditions (920 - 928~nm), trapping with SIBA was still favorable due to the dynamic reconfiguration of the trap that allowed the particle to be trapped in low intra-cavity intensity.

An interesting observation in Figures~\ref{Fig5}a,b is the fluctuations in the trapping stiffness with increasing incident power and the deviation from the linear increase that is theoretically predicted. This is attributed to the effect of additional forces, such as thermal and fluid-related forces, that may play a significant role in the trapping mechanism. 

\paragraph{Thermal forces:}
The height of the microwell containing the gold particles was 120~$\mu$m.  Photo-induced heating resulting from intrinsic losses of the metallic nanostructures generates thermal convection currents and thermophoresis that could influence the trapping process~\cite{Kotsifaki2022TheRO}. Specifically, as the laser trapping intensity increased, the optically generated temperature gradients could induce fluid flows that quickly delivered the nanoparticles above the nanostructures. To maximize particle-trapping probabilities, the nanoparticle should be delivered with low fluid velocity to the hotspot where it would  be trapped under the short-range plasmonic fields. However, when the height of the solution's microwell is larger than 20~$\mu$m the fluid velocity is large, thereby preventing stable trapping~\cite{Xu2019AlldielectricNF, Braibanti}. In addition, thermophoresis induced by the temperature gradients could alter thermophoretic migration of particles~\cite{Braibanti}. For temperatures higher than a critical temperature, the Soret coefficient sign changes~\cite{WurgerSoret} and thermophoresis repels  particles away from a high-temperature location~\cite{Kotsifaki2022TheRO}, hindering trapping stability. 

\paragraph{Fluid-related forces:}
An important factor that may impact the trapping performance is the role of the Triton X-100 in the liquid solution. It has been reported that a small amount of non-ionic surfactant generates thermodiffusion with opposite sign from the optical force leading to fluctuations of the particle's position inside the trap and a lower trap stiffness magnitude~\cite{Jiang2020QuantifyingTR}. 

\paragraph{Electrostatic and scattering forces:}
An additional effect may be present due to electrostatic repulsive forces as the gold particle (-24~mV zeta potential) approaches the gold surface, with both being negatively charged~\cite{RodrguezSevilla2018OpticalFA}. This can result in a decreased measured trapping stiffness since our evaluation method is based on the motion of the particle as it enters the trap. Finally, upon illumination of the gold structure, light scattering may contribute to the destabilisation of the particle and the deviation from the linear increase of the trapping stiffness.

As the combination of the effects above strongly depend on the particle and solution properties, as well as local temperature effects, the trapping landscape can get quite complicated and an optimized trapping scheme is required that would take advantage of all the positive phenomena and minimise the negative ones. However, this is beyond the scope of this paper and it will constitute a future work.

\section{Conclusions}
Trapping of gold nanoparticles was performed via metamaterial plasmonic tweezers. The investigated trapping conditions were for on-resonant and blue-detuned wavelengths to enable the SIBA effect. Very high experimental trapping stiffness values were observed, with on-resonant trapping exhibiting the highest value at 2.92~$\pm$~0.42~fN/(nm$\cdot$mW/$\mu$m$^{2}$) for an excitation laser intensity as low as 0.61~mW/$\mu$m$^2$. For the off-resonant case, the highest value was  4.18~$\pm$~0.2~fN/(nm$\cdot$mW/$\mu$m$^{2}$) but for an excitation intensity about 62\% higher compared with the on-resonant case. The experimental results had the characteristic feature of SIBA-assisted trapping, where the maximum trapping stiffness is exhibited for the lowest excitation intensity. For the SIBA-resonant wavelength of 930~nm, two distinct regimes were observed; the low-intensity regime where the SIBA was the main mechanism contributing to the trap and the high-intensity regime where the SIBA vanished and the optical forces were dominating in the trapping mechanism, in agreement with theory. Blue-detuned trapping exhibited a more complex behavior that deviates from the linear increase of the trapping stiffness with increased excitation intensity. Additional forces, such as thermal, fluid, and electrostatic forces, may have contributed to this deviation. Extensive simulations were also performed and revealed the existence of two groups of hotspots in each unit cell of the metamaterial array. With varying excitation wavelength a different ratio of \textit{E}-field intensities could be excited at each hotspot, thus resulting in traps of different strengths enabling the sorting of particles based on their size, shape, and refractive index. This could prove extremely useful in biological applications for sorting, for example, healthy from damaged distorted cells, and investigating their morphological, kinetic, and other characteristics.

\newpage
\subsection{Associated Content}
\paragraph{Supporting Information}
\noindent Details on the sample fabrication, solution preparation and simulations can be found in the Methods section of the Supplementary Information. Additional simulation figures and analytical calculations of the cavity shift and the optomechanical coupling constant are also included.\\
S1: Methods\\
S2: Experimental and simulated structure\\
S3: Cavity shift and optomechanical coupling constant\\
S4: Simulated optical forces\\
S5: Simulated forces and potentials at the two hotspots\\

\subsection{Authors Information}

\textbf{Theodoros D. Bouloumis} $^{}$\orcidA{}:0000-0002-5264-7338,\\
Light – Matter Interactions for Quantum Technologies Unit, Okinawa Institute of Science and Technology Graduate University, 1919-1 Tancha, Onna-son, Okinawa, 904-0495, Japan, theodoros.bouloumis@oist.jp\\

\noindent \textbf{Domna G. Kotsifaki} $^{}$\orcidB{}:0000-0002-2023-8345,\\
Natural and Applied Sciences, Duke Kunshan University, No. 8 Duke Avenue, Kunshan, Jiangsu Province, 215316, China,\\
Light – Matter Interactions for Quantum Technologies Unit, Okinawa Institute of Science and Technology Graduate University, 1919-1 Tancha, Onna-son, Okinawa, 904-0495, Japan,\\
domna.kotsifaki@dukekunshan.edu.cn\\

\noindent \textbf{S\'ile Nic Chormaic} $^{}$\orcidC{}:0000-0003-4276-2014, \\
Light – Matter Interactions for Quantum Technologies Unit, Okinawa Institute of Science and Technology Graduate University, 1919-1 Tancha, Onna-son, Okinawa, 904-0495, Japan, sile.nicchormaic@oist.jp

\subsection{Notes}
TDB and DGK conceived the idea of this work, DGK prepared the solutions, TDB performed the experiments, simulations, and analyzed the data. SNC supervised all stages of the work.  All authors contributed to writing the paper.\\

\noindent The authors declare no competing financial interests.

\begin{acknowledgement}

The authors thank M. Ozer for technical support, J. Keloth for useful discussions, and N. Kokkinidis for his input regarding simulations. They also acknowledge P. Puchenkov and J. Moren from the Scientific Computing and Data Analysis Section, and the Engineering Section of the Research Support Division at OIST. This work was partially supported by the Okinawa Institute of Science and Technology Graduate University. TDB acknowledges funding from the Japan Society for the Promotion of Science (JSPS) KAKENHI Grants-in-Aid (Grant No. 22J10196) and from the International Society for Optics and Photonics (SPIE) for the Optics and Photonics Education Scholarship 2021.

\end{acknowledgement}

\newpage
\begin{suppinfo}

\setcounter{section}{0}
    \renewcommand{\thesection}{S\arabic{section}}
    \setcounter{equation}{0}
    \renewcommand{\theequation}{S\arabic{equation}}
    \setcounter{figure}{0}
    \renewcommand{\thefigure}{S\arabic{figure}}

\subsection{S1 Methods}

        The simulations were performed on the software \textit{COMSOL Multuphysics} with the \textit{Wave Optics} module, which is based on the finite element method (FEM). The geometry of the simulated structures was built entirely on \textit{COMSOL} with the dimensions matching the experimental dimensions measured via SEM (Fig.~\ref{SI_structure}a). The mesh was set to finer - custom with minimum size 1.2~nm, maximum size 44.3~nm and resolution of narrow regions at 0.7~nm. The Floquet periodicity was used with a periodic port from which we illuminated with a plane wave. Initially, we performed a frequency domain study in order to simulate the reflection, transmission, and absorption spectra which gave us the theoretical resonance wavelength of the plasmonic metamaterial cavity. Next, we introduced a gold nanoparticle with 20~nm diameter, surrounded by a 2.5~nm thick water shell and scanned its position along the three directions to calculate the optical forces exerted on the nanoparticle. For the force calculation, the Maxwell stress tensor~\cite{griffiths_2017} was used with surface integration around the water shell.
    
        The metamaterial structures were fabricated on a 50~nm gold thin film (PHASIS, Geneva, BioNano) using focused ion beam (FIB) milling. The FIB system was a FIB-FEI Helios Nanolab G3UC and the structures were etched at 80~nm depth to ensure that the structures were cut all the way through the gold film uniformly. The voltage and beam current used for etching were 30~kV and 2~pA, respectively and the volume per dose was 0.27~$\mu$m$^3$/nC. The scanning electron microscopy (SEM) images were acquired with the same system. Each metamaterial array consisted of 15~$\times$~15 units with a nominal period of 400~nm. The geometrical characteristics of the metamaterial are shown in Figure~\ref{SI_structure}a. After fabrication, the device underwent oxygen plasma treatment for 3 minutes to remove gallium residue that was deposited on it during the milling process~\cite{SuKo_Ga}. The transmission spectra were obtained with a microspectrophotometer (CRAIC, 20/30 PV) with the device immersed in a miliQ water environment. 
        
        For the particle solution, we prepared an initial mixture of heavy water (D$_2$O, Sigma-Aldrich) and phosphate buffered saline (PBS) 10x buffer (pH~=~7.4, Invitrogen) with ration 9/1, and dispersed  20~nm diameter gold particles (Sigma-Aldrich No:753610) in it with a concentration of 0.3\%~v/v. Surfactant Triton X-100 (Sigma-Aldrich) with a concentration of 1.2\%~v/v was also added to avoid the aggregation of particles. The solution was sonicated for 10 minutes and then 8.7~$\mu$L of the final solution was placed inside a seal spacer with 9~mm diameter and 120~$\mu$m thickness (Invitrogen) on the gold film where the nanostructures were fabricated. Finally, the sample was sealed from the top with a cover glass, forming a microwell.
        
        The sample was mounted on a custom-made inverted microscope setup. The excitation laser was incident from the bottom side of the sample (SiO$_2$ glass substrate) and focused on the structures through a high numerical aperture (NA~=1.25) 100x oil immersion objective lens (Nikon MRP71900, E Plan). The transmitted light was collected from the other side through a second objective lens (Nikon CF Plan BD DIC 20x) and detected on an avalanche photodetector (APD430A/M Thorlabs) for real-time monitoring of the signal. The APD was connected to a data acquisition module (DAQ USB-6363, NI) and interfaced using LabVIEW code to acquire measurements at a 100~kHz rate.

\subsection{S2 Experimental and simulated structure}

\begin{figure}
    \centering \captionsetup{justification=raggedright,singlelinecheck=false}
    \begin{subfigure}[t]{0.3\textwidth}
        \centering
        \caption{}
        \includegraphics[width=\textwidth]{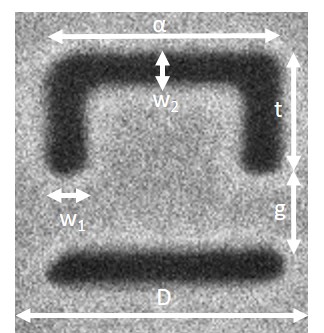}
        \label{SI_metanunit}
      \end{subfigure}
    \hspace{2cm}
    \begin{subfigure}[t]{0.38\textwidth}
        \centering
        \caption{}
        \includegraphics[width=\textwidth]{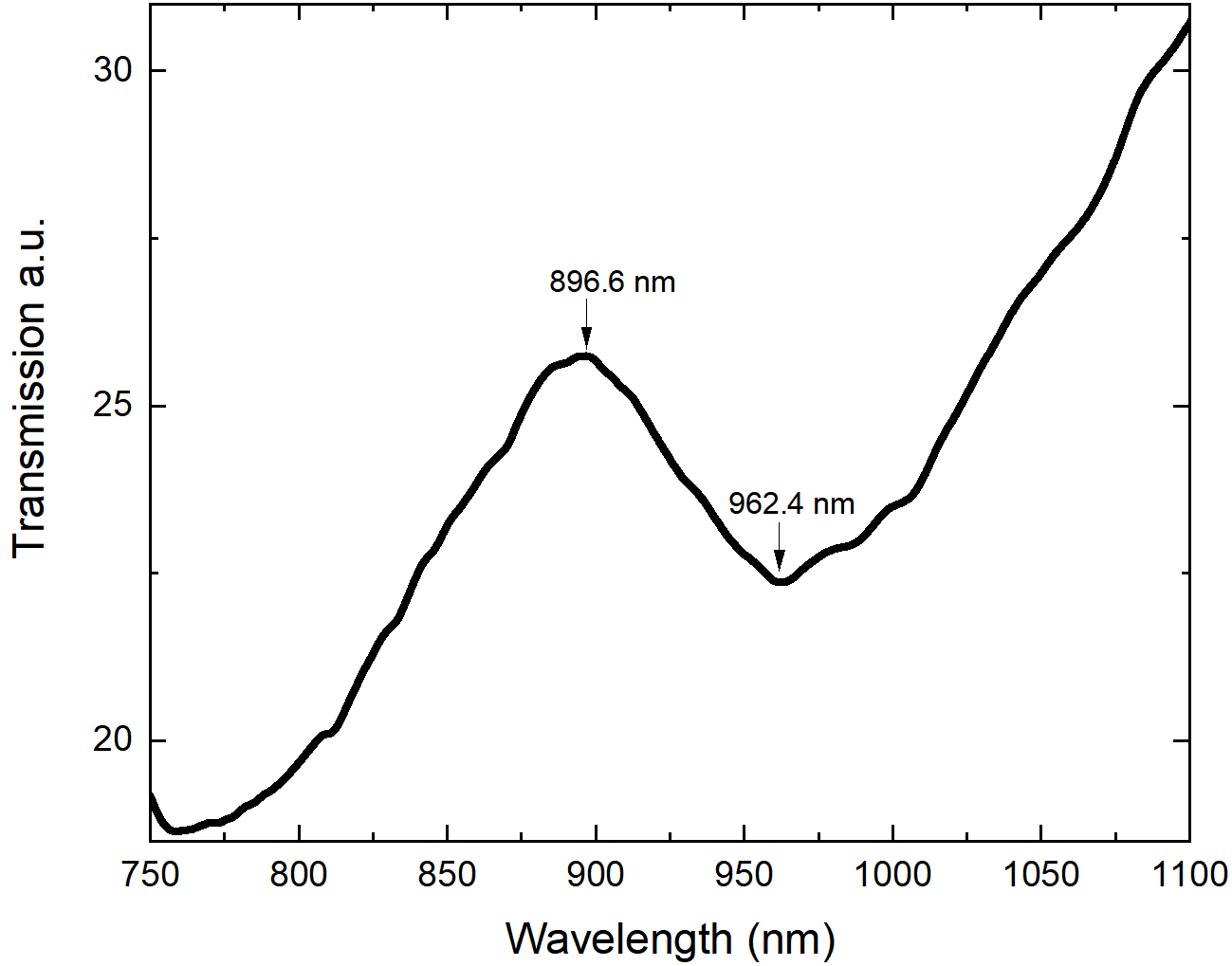}
        \label{SI_transmission}
    \end{subfigure}
    \hspace{1cm}
    \begin{subfigure}[t]{0.5\textwidth}
        \centering
        \caption{}
        \includegraphics[width=\textwidth]{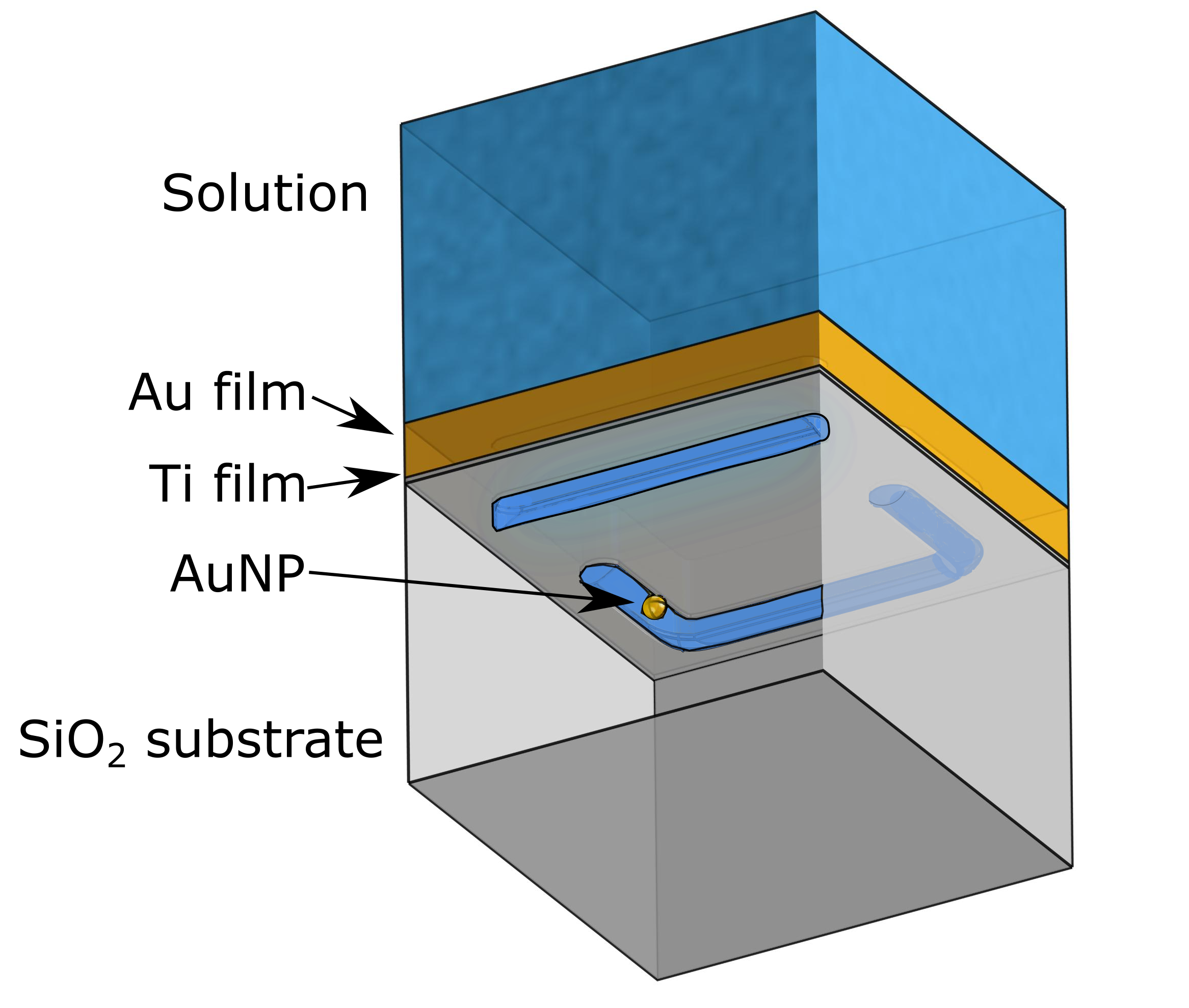}
        \label{SI_design}
    \end{subfigure}
    \hspace{0.5cm}
     \begin{subfigure}[t]{0.4\textwidth}
         \centering
         \caption{}
         \includegraphics[width=\textwidth]{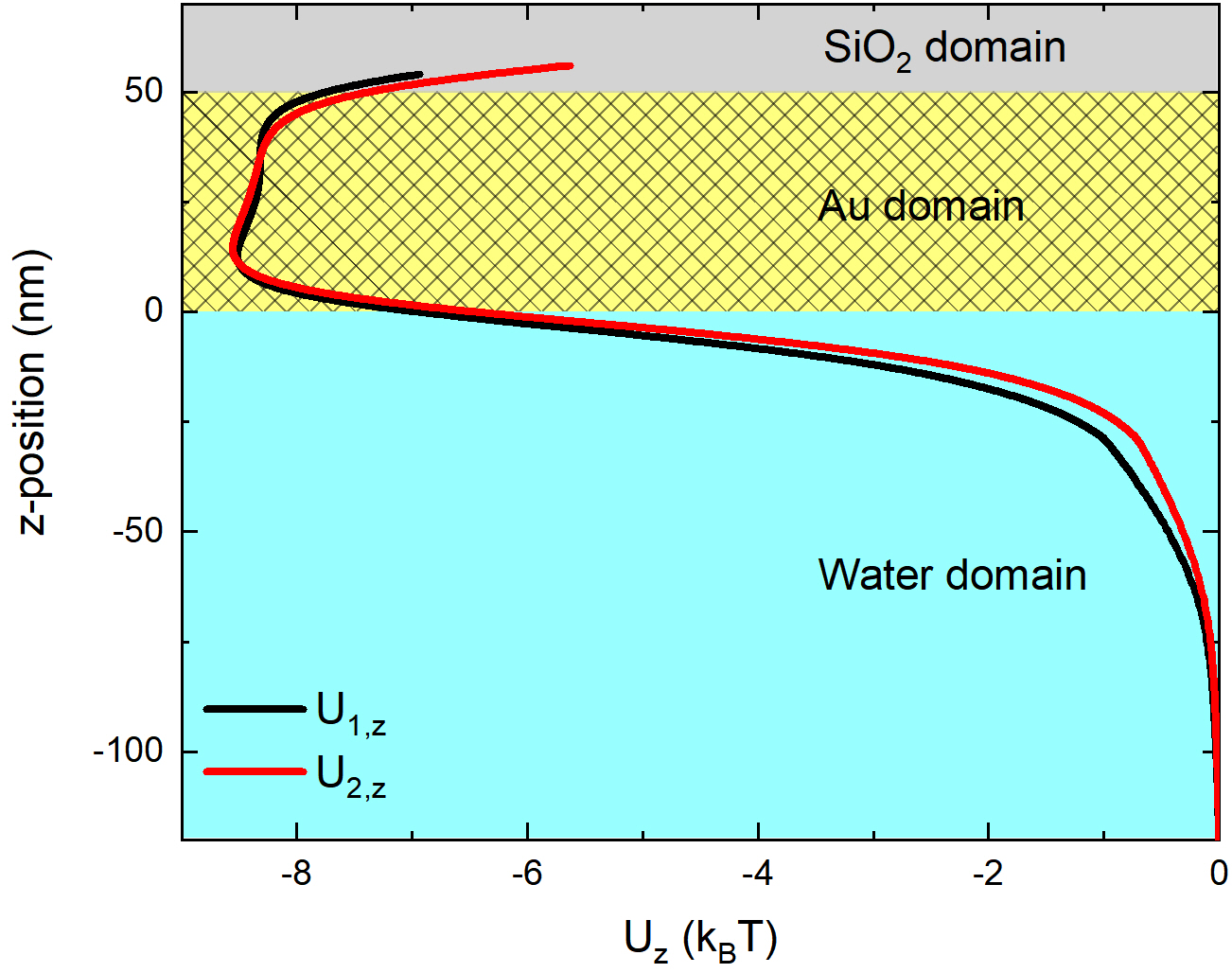}
         \label{SI_potential}
     \end{subfigure}
     \captionsetup{justification=justified}
    \caption{(a) SEM image of a unit cell of the metamaterial structure. The dimensions are: $w_1$~=~39.83~$\pm$~0.77~nm, $w_2$~=~33.33~$\pm$~0.66~nm, $\alpha$~=~307.55~$\pm$~0.55~nm, $t$~=~157.23~$\pm$~0.54~nm, $g$~=~108.80~$\pm$~0.91~nm, $D$~=~416.14~$\pm$~1.00~nm. (b) Experimental transmission spectrum. (c) The simulated design based on the experimental sample. The etching of the sample goes about 20-30~nm into the glass substrate to ensure uniform cut throughout the gold film. (d) Simulated potentials along the z-axis for the two hotspots and for an excitation wavelength of 930~nm.}
    \label{SI_structure}

\end{figure}

The experimental dimensions of the unit array are shown in Figure~\ref{SI_metanunit} and defined as $w$, the slot width, $\alpha$, the length of the long arm, $t$, the length of the short arm, $g$, the separation between the "C" and "I" parts of the structure, and $D$ the period of the array. 
\\

\noindent From the experimental transmission spectrum (Fig.~\ref{SI_transmission}), we can calculate the absorption peak wavelength, which indicates the Fano resonance, from~\cite{Fan2018AchievingAH}

\begin{equation}
     \lambda_0 = \frac{2\lambda_{peak} \lambda_{dip}}{\lambda_{peak} + \lambda_{dip}} = 928.3~\text{nm},
\end{equation}
where $\lambda_{peak}$ and $\lambda_{dip}$ correspond to the wavelengths of the Fano peak indicated in Figure~\ref{SI_transmission}.

\newpage
\subsection{S3 Cavity shift and optomechanical coupling constant}

\paragraph{Cavity shift:}

Following work by Mestres et al.~\cite{Mestres2016Unravelling}, a trapped nanoparticle in a plasmonic cavity induces a plasmonic resonance shift given by

\begin{equation}
 \delta \omega (r_p)=\omega_c  \frac{\alpha_d}{2V_m\epsilon_0} f(r_p),  
 \label{cavity_shift}
\end{equation}
where $\omega_c$ is the empty cavity resonance, $\alpha_d$ the effective polarizability of the particle, $V_m $ mode volume of the cavity, and $f(r_p)$ the cavity intensity profile. 

The cavity resonance according to simulations is at $\lambda_c~=~928$nm or $\omega_c~=~2~\cdot~10^{15}$$~rad/sec$ and the mode volume of the cavity is given with a rough approximation by~\cite{Tanaka2010}

\begin{equation}
    V_m = 2(\alpha + t - w_2)\frac{w_1 + w_2}{2} \cdot h
    \label{mode_volume}
\end{equation}
where $\alpha, t, w_1, w_2$ are the geometrical characteristics noted in Figure~\ref{SI_metanunit} and $h$ is the thickness of the gold film, in our case 50~nm. The average value of $w_1$ and $w_2$ is assumed as the general slot width for the equation. The calculated value is $V_m~=~1.58~\cdot~10^{-3}~\mu m^3$.

For the polarizability of the particle, the radiative reaction correction to the Clausius - Mossotti relation was taken into account. For a particle with radius $r~\ll~\lambda$ the effective polarizability, $\alpha_d$, is given by~\cite{jones_marago_volpe_2015}

\begin{equation}
    \alpha_d = \dfrac{\alpha_0}{1 - \frac{\epsilon_r - 1}{\epsilon_r + 2} [(k_0r)^2 + \frac{2i}{3}(k_0r)^3]},
    \label{effective_polarisability}
\end{equation}
where $\alpha_0$ is the polarizability of the particle given by the Clausius - Mossoti relation

\begin{equation}
    \alpha_0=4\pi r^3 \epsilon_0 \dfrac{\epsilon_r -1}{\epsilon_r+2},
    \label{polarizability}
\end{equation}
$k_0$ is the wavenumber, $\epsilon_0$ the vacuum dielectric permittivity, and $\epsilon_r$ is the particle's dielectric permittivity taken from ~\cite{PhysRevX.10.011071} to be $\epsilon_r~=~-35.618 + 1.7121i$.

The intensity profile of the cavity, $f(r_p)$, was taken equal to 1, which is its maximum normalized value for maximum intensity obtained when exciting the cavity with the absorption resonant wavelength of 928~nm~\cite{Mestres2016Unravelling}.

Substituting the values of Equations~\ref{mode_volume} and \ref{effective_polarisability} in Equation~\ref{cavity_shift}, the resulting cavity shift expressed as a wavelength shift is $\delta\lambda~=~4.06$~nm.

\paragraph{Optomechanical coupling constant:}

The optomechanical coupling constant is given approximately by~\cite{Mestres2016Unravelling} 

\begin{equation}
    G~\approx~\frac{\delta\omega}{\delta x},
    \label{optomechanical}
\end{equation}
where $\delta\omega$ is the cavity shift calculated above equal to 2$\pi\cdot$1.41~THz, and $\delta x$ is the displacement of the particle from the equilibrium position assumed to be equal to the particle's diameter as a convention. Thus, G~=~2$\pi\cdot$~71~GHz/nm.

\newpage
\subsection{S4 Simulated optical forces and potentials}

\begin{figure}
     \centering \captionsetup{justification=raggedright,singlelinecheck=false}
     \begin{subfigure}[t]{0.45\textwidth}
         \centering
         \caption{}
         \includegraphics[width=\textwidth]{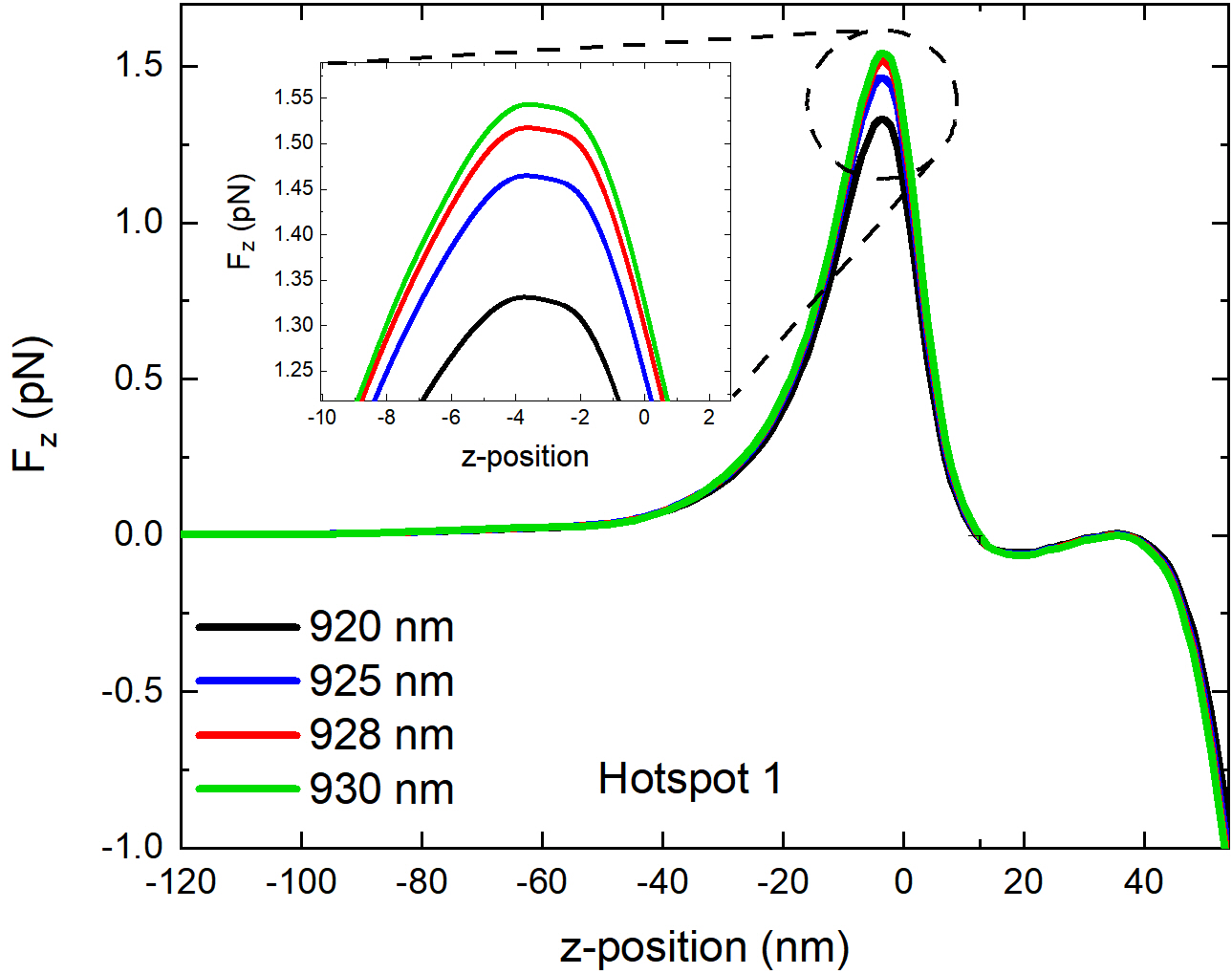}
         \label{FigF1z}
     \end{subfigure}
     \hfill
     \begin{subfigure}[t]{0.45\textwidth}
         \centering
         \caption{}
         \includegraphics[trim={0 0 0 0.2cm}, clip, width=\textwidth]{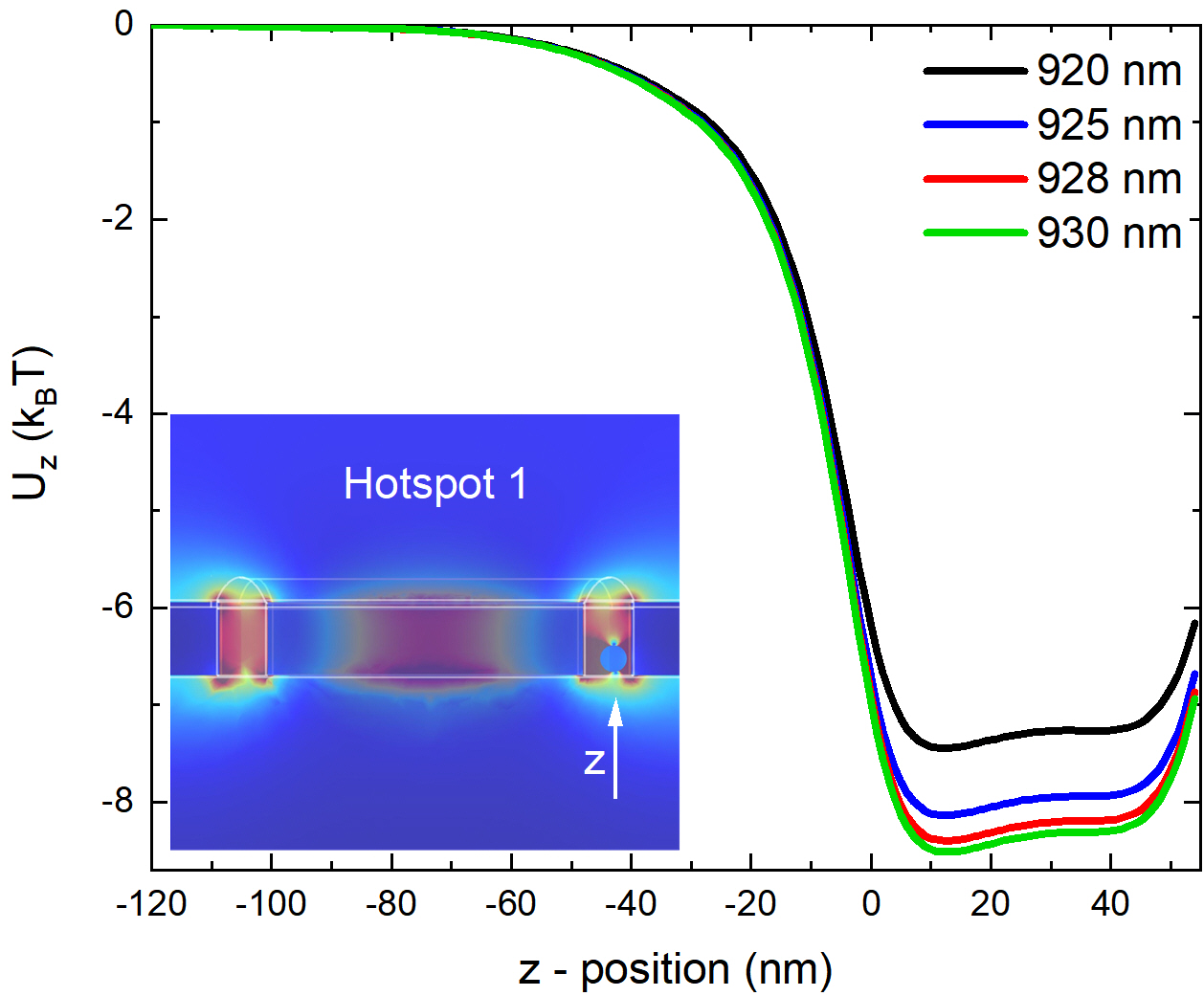}
         \label{FigU1z}
     \end{subfigure}
     \vspace{10pt} 
     \begin{subfigure}[t]{0.45\textwidth}
         \centering
         \caption{}
         \includegraphics[width=\textwidth]{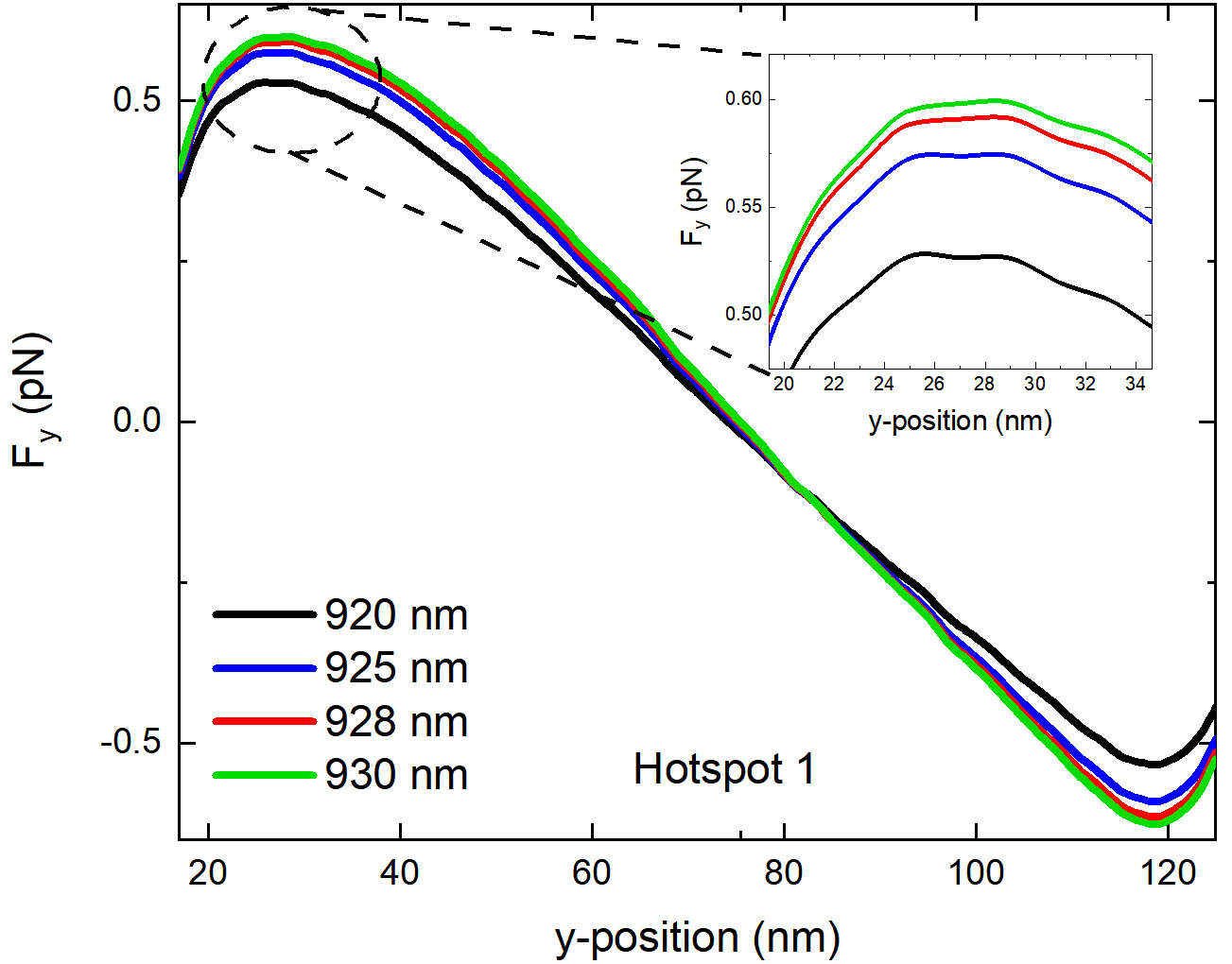}
         \label{FigF1y}
     \end{subfigure}
     \hfill
     \begin{subfigure}[t]{0.45\textwidth}
         \centering
         \caption{}
         \includegraphics[width=\textwidth]{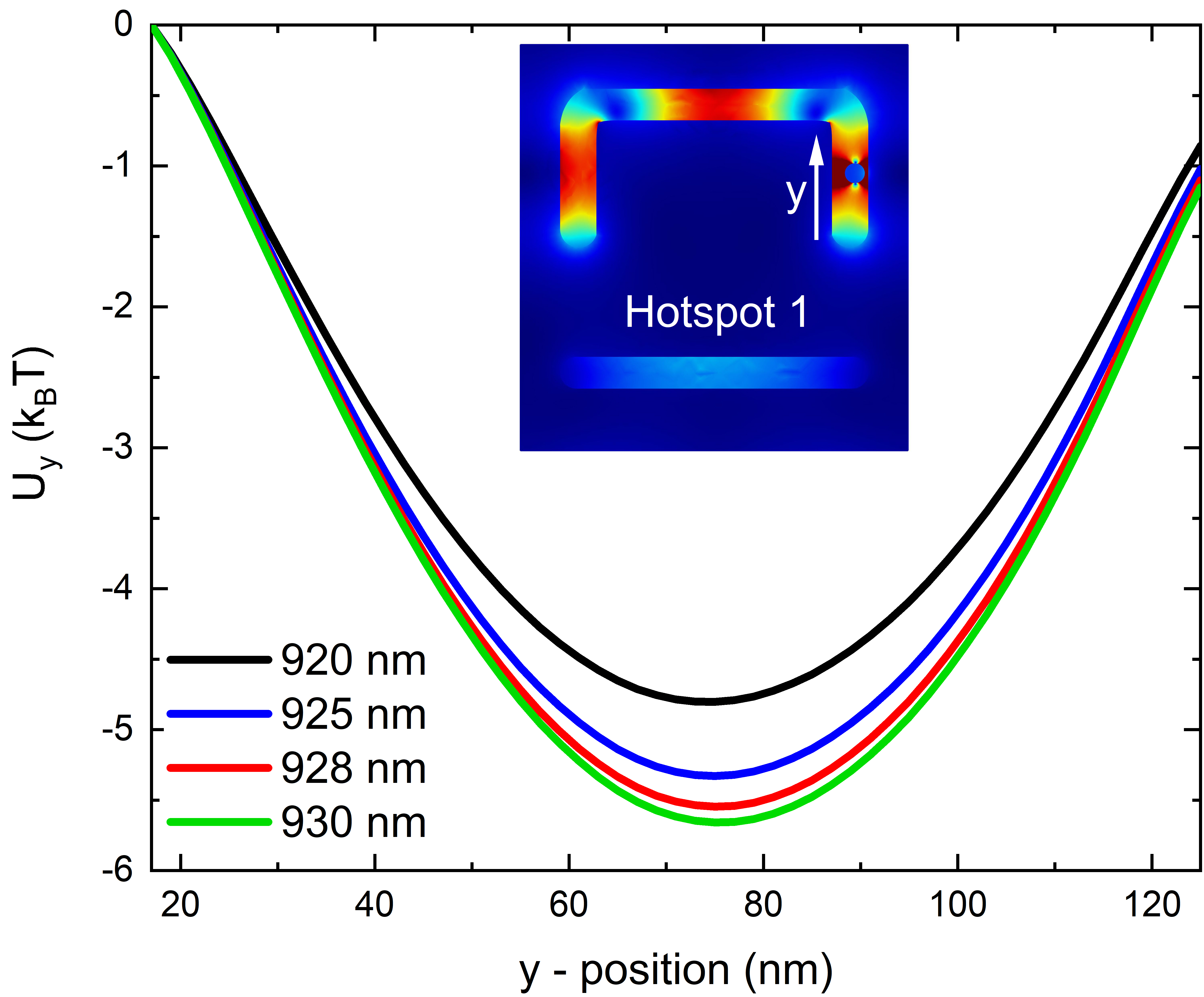}
         \label{FigU1y}
     \end{subfigure}
        \captionsetup{justification=justified}
        \caption{Simulated optical forces (a), (c) and their corresponding potentials (b),(d) experienced by a gold nanoparticle with 20~nm diameter around Hotspot 1. The calculations were performed along the (a),(b) $z$- and (c),(d) $y$-axis for excitation wavelengths 920~nm, 925~nm, 928~nm, and 930 nm.}
        \label{SI_F_U_hotspot1}
\end{figure}

In Figure~\ref{SI_F_U_hotspot1}, the optical forces and potentials are plotted along the $z$ and $y$ axes, for a particle around Hotspot 1 and for excitation wavelengths 920~nm, 925~nm, 928~nm, and 930~nm. The potential depth and optical forces exerted on the particle increase as we scan the excitation wavelength from 920~nm to 930~nm, with 930~nm creating the strongest force. The maximum force along the $z$-axis for the case of 930~nm is around 1.5~pN, normalised for 1~mW excitation power. For all the wavelengths, the maximum force is exerted at the position $z=-3$~nm which is inside the water domain 3~nm away from the surface of the gold film, whereas the equilibrium position is at $z=12.6$~nm, inside the gold domain which is etched and filled with water (Fig.~\ref{SI_design}). The force along the $x$-axis, although not presented here, was calculated and it was always pushing the particle toward the external wall of the cavity.

\begin{figure}[ht]
    \centering \captionsetup{justification=raggedright,singlelinecheck=false}
     \begin{subfigure}[t]{0.4\textwidth}
         \centering
         \caption{}
         \includegraphics[width=\textwidth]{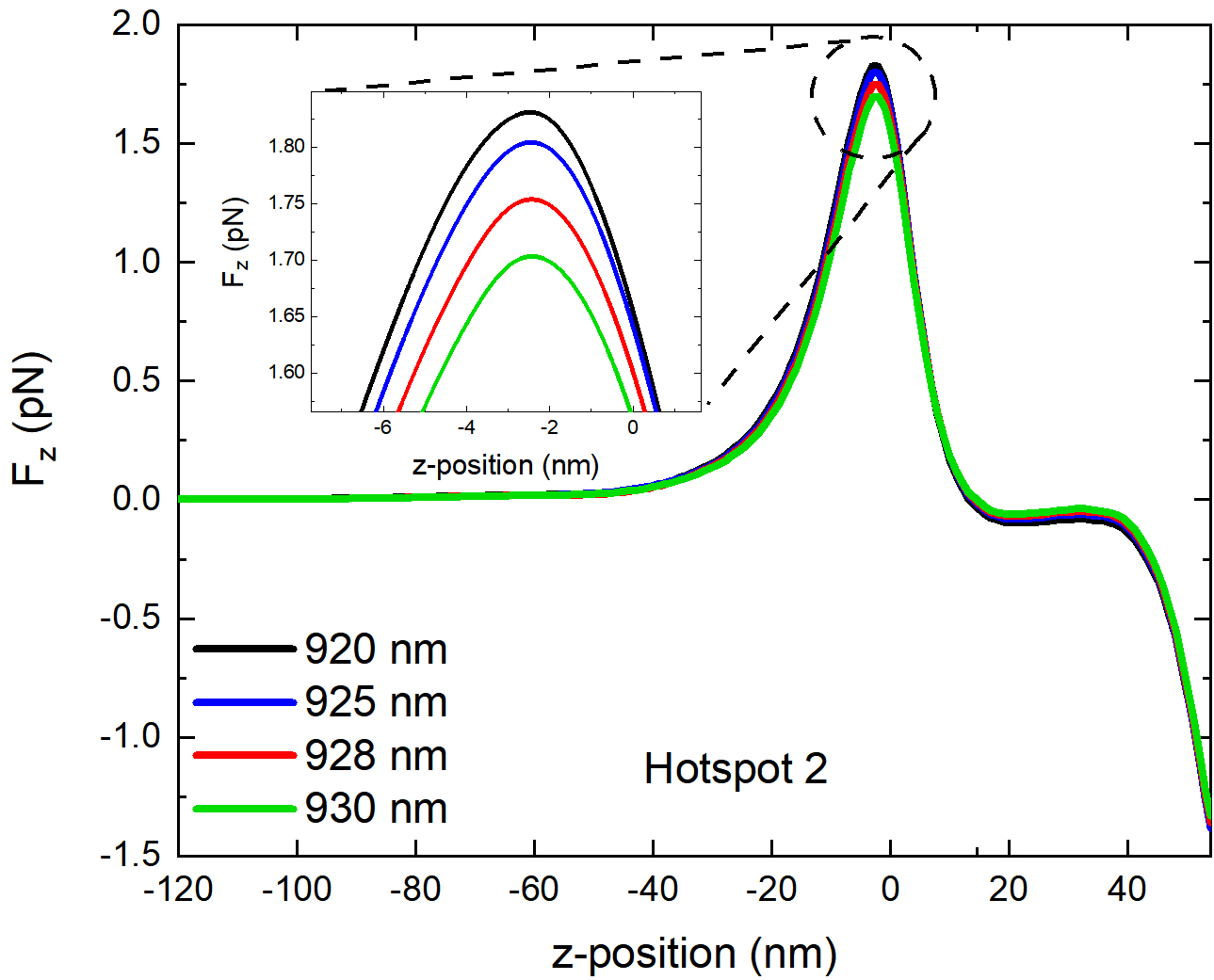}
         \label{FigF2z}
     \end{subfigure}
    \hspace{2cm}
     \begin{subfigure}[t]{0.4\textwidth}
         \centering
         \caption{}
         \includegraphics[width=\textwidth]{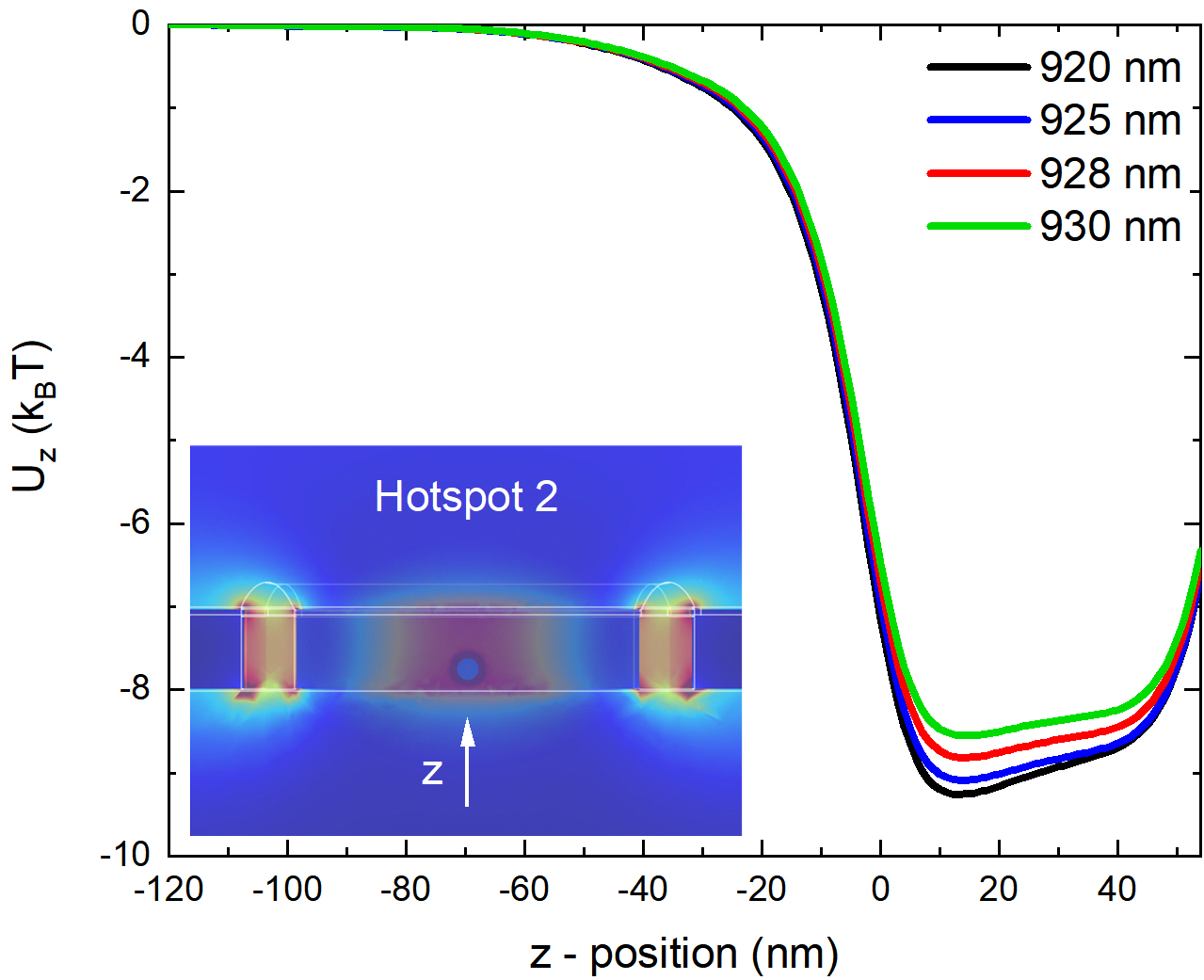}
         \label{FigU2z}
     \end{subfigure}
    \hspace{2cm}
     \begin{subfigure}[t]{0.4\textwidth}
         \centering
         \caption{}
         \includegraphics[width=\textwidth]{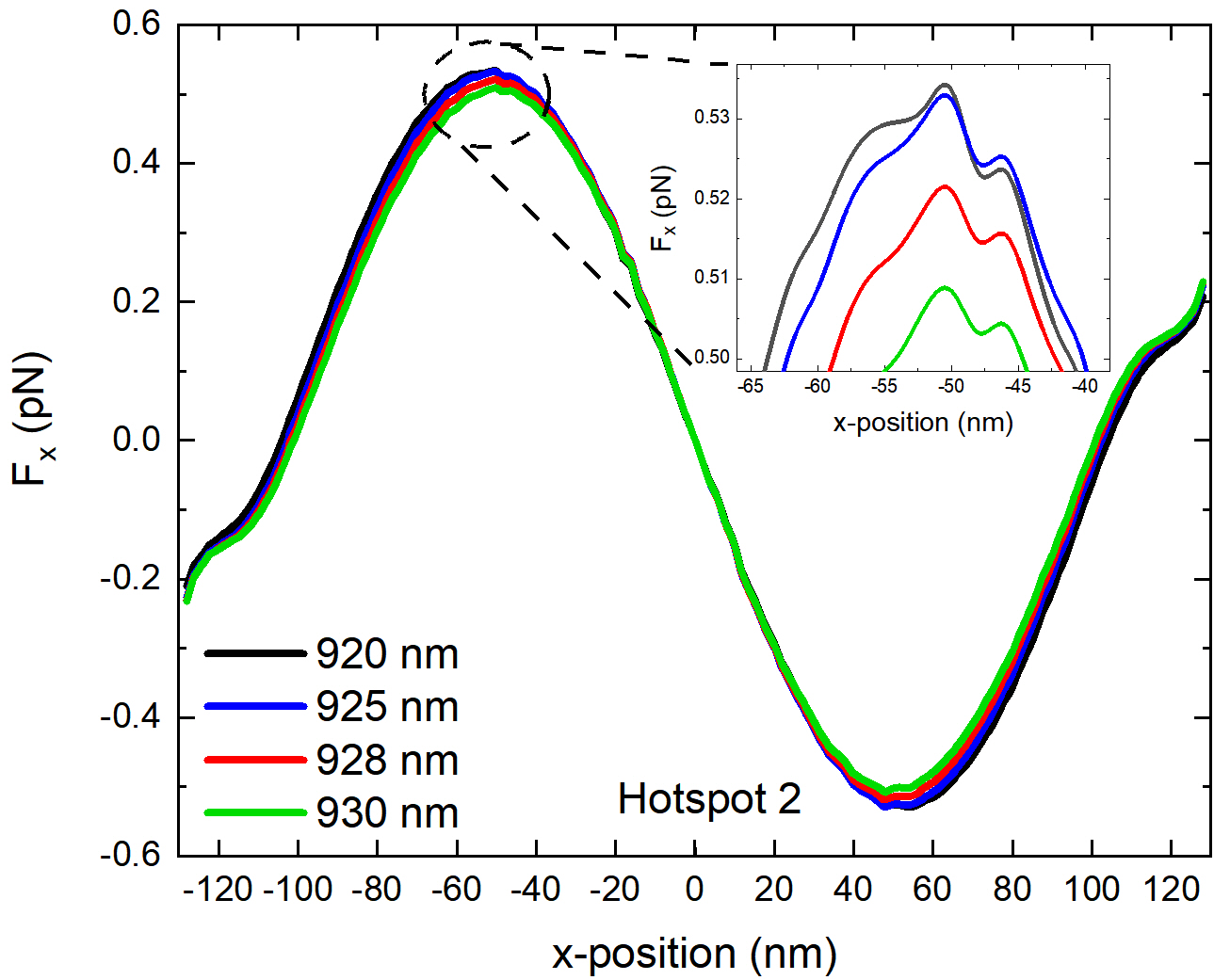}
         \label{FigF2x}
     \end{subfigure}
    \hspace{2cm}
        \begin{subfigure}[t]{0.4\textwidth}
         \centering
         \caption{}
         \includegraphics[width=\textwidth]{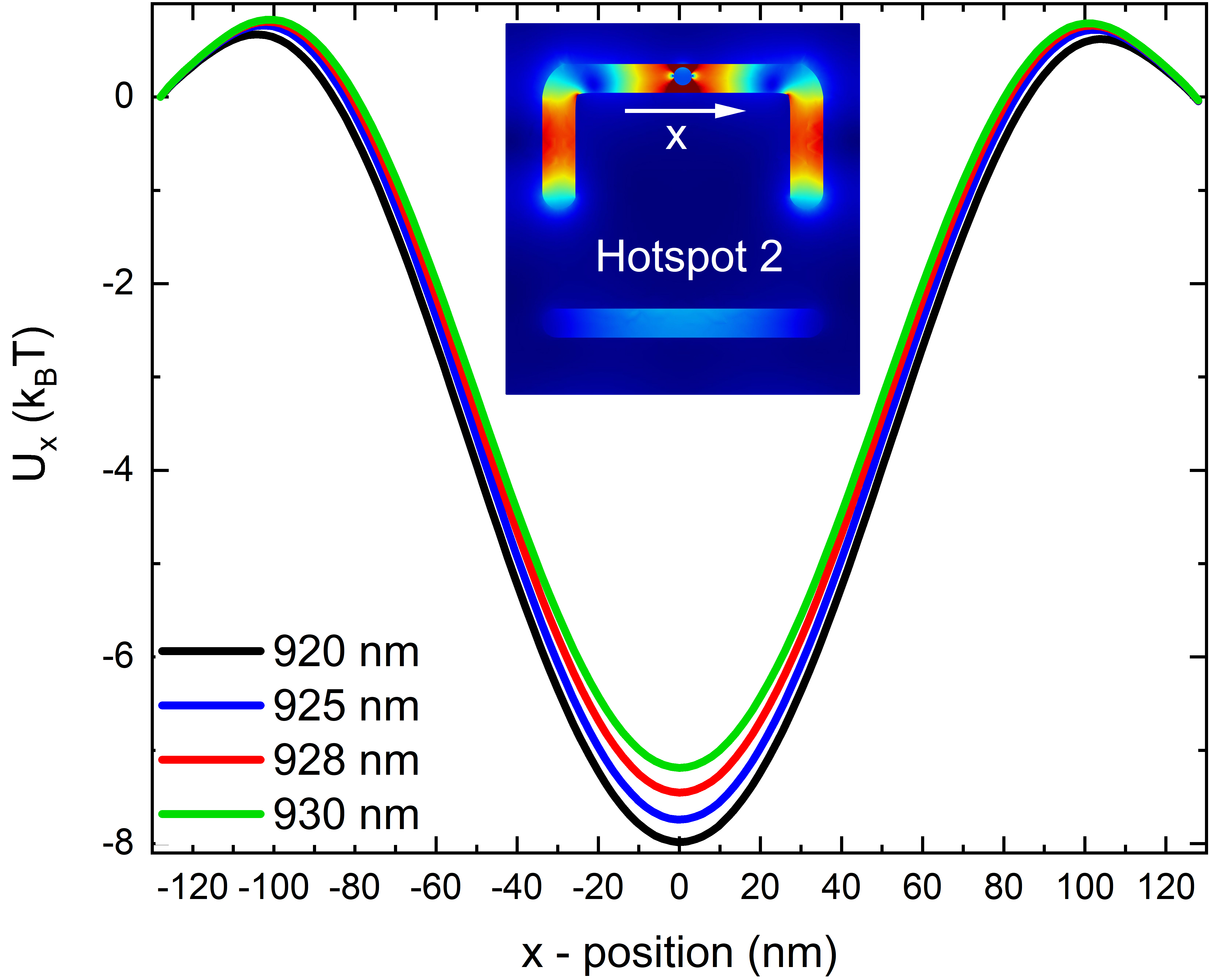}
         \label{FigU2x}
     \end{subfigure}
     \captionsetup{justification=justified}
     \caption{Simulated optical forces (a), (c) and their corresponding potentials (b),(d) experienced by a gold nanoparticle with 20~nm diameter around Hotspot 2. The calculations were performed along the (a),(b) $z$- and (c),(d) $y$-axis for excitation wavelengths 920~nm, 925~nm, 928~nm, and 930 nm.}
     \label{SI_F_U_hotspot2}

\end{figure}

Similarly, the forces and potentials are plotted for Hotspot 2 in Figure~\ref{SI_F_U_hotspot2}; however, in this case, the potential depths and optical forces are decreasing as we scan from 920~nm to 930~nm which is the opposite situation compared to Hotspot 1. Additionally, for all wavelengths, the forces at Hotspot 2 are slightly higher than those at Hotspot 1, with the maximum force and equilibrium positions at $z=-2.5$~nm and $z=13.5$~nm, respectively. Note that the scale bar in all the E-field insets is the same as the one in Figure~\ref{SI_hotspots}.

\subsection{S5 Simulated forces and potentials at the two hotspots}

\begin{figure}[ht]
    \centering \captionsetup{justification=raggedright,singlelinecheck=false}
     \begin{subfigure}[t]{0.46\textwidth}
         \centering
         \caption{}
         \includegraphics[width=\textwidth]{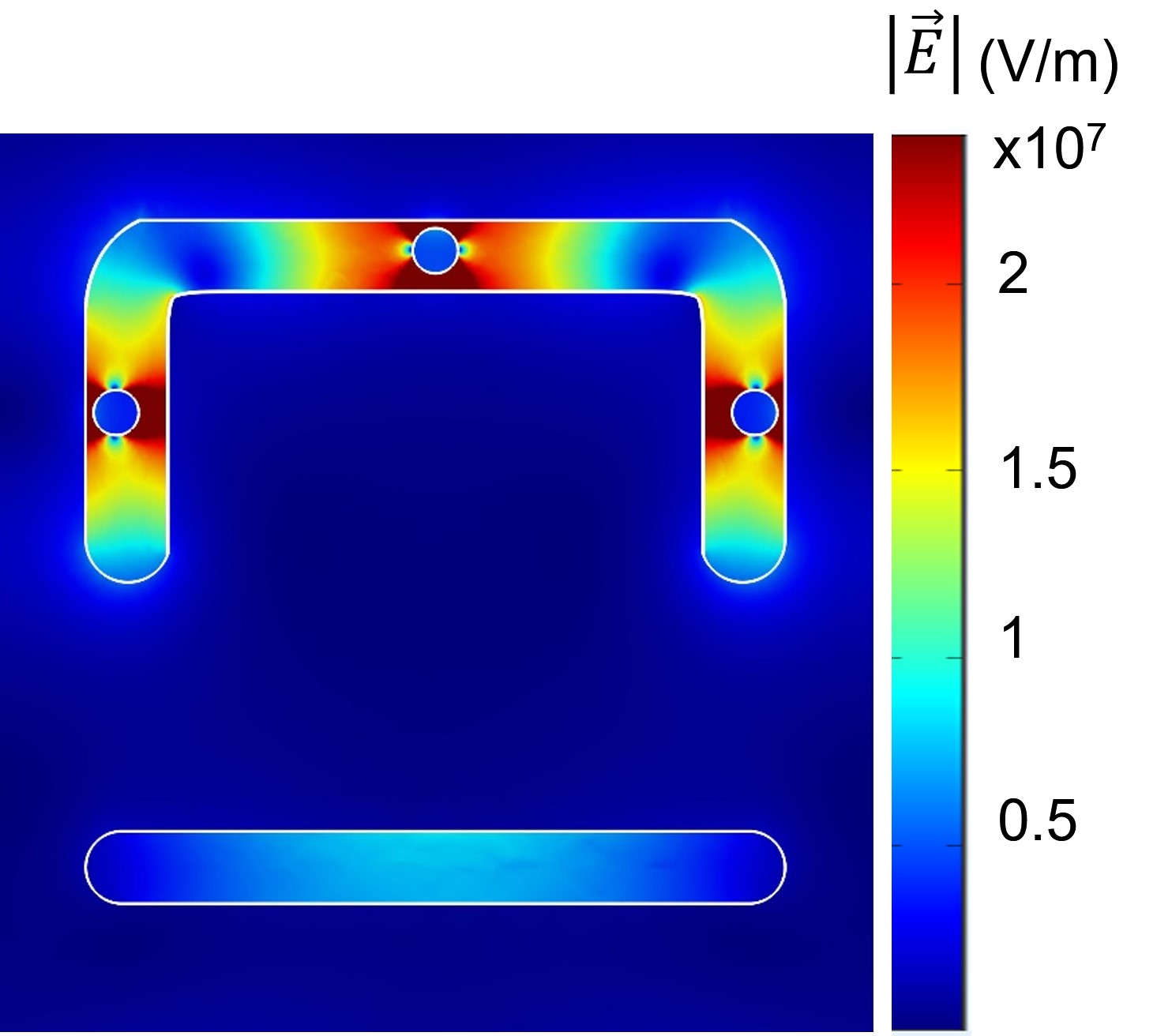}
         \label{SI_hotspots}
     \end{subfigure}
    \hspace{1cm}
     \begin{subfigure}[t]{0.41\textwidth}
         \centering
         \caption{}
         \includegraphics[width=\textwidth]{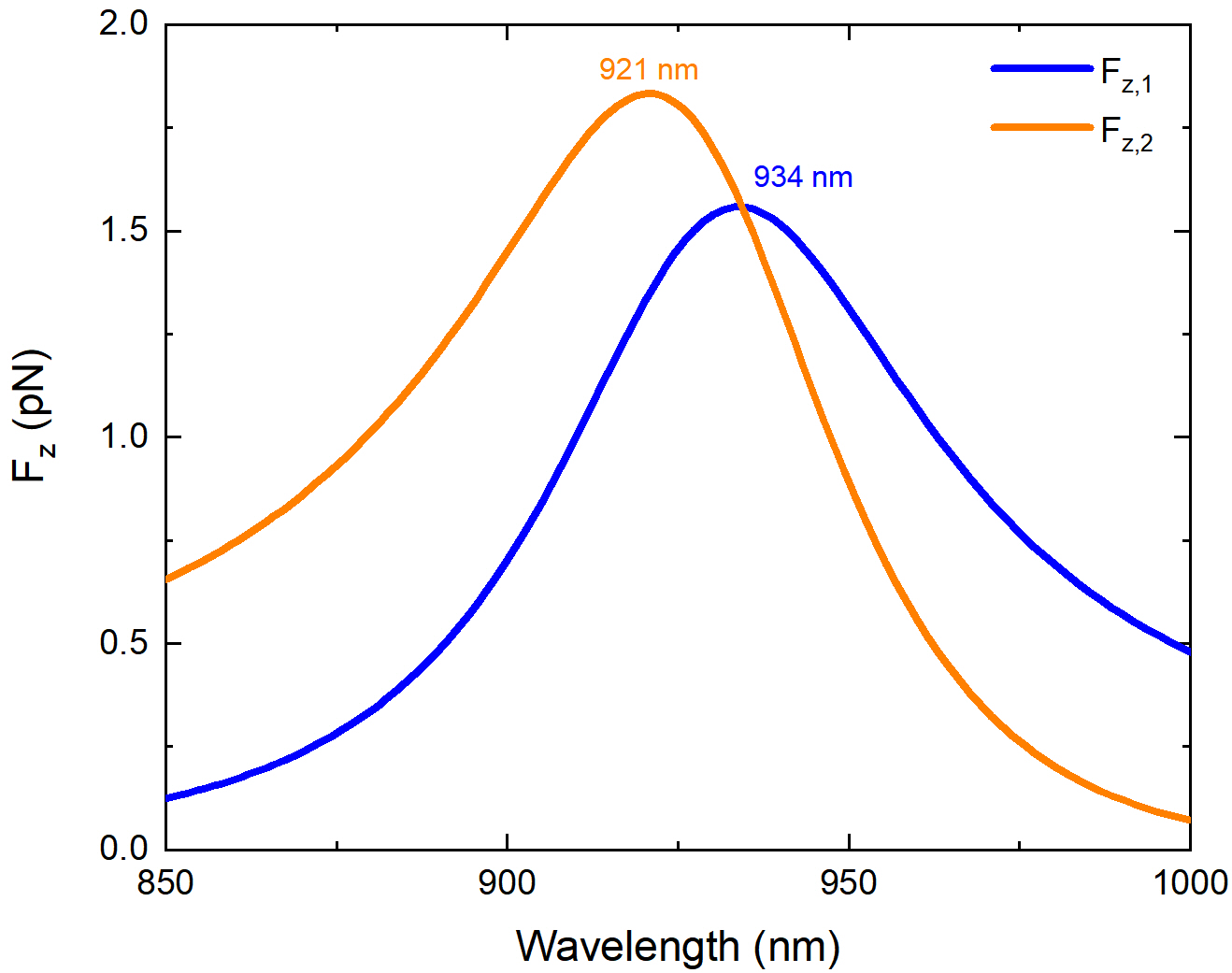}
         \label{SI_F-lamda}
     \end{subfigure}
    \captionsetup{justification=justified}
     \caption{(a) Simulated electric field at a unit cell with particles trapped at the hotspots. The narrower slit creates better particle confinement and higher electric field enhancement. (b) Simulated forces acting on a gold nanoparticle at the equilibrium position of Hotspot 1 (blue line) and Hotspot 2 (orange line). }
     \label{SI_simulations}
\end{figure}

\end{suppinfo}

\newpage
\bibliography{Manuscript}

\providecommand{\latin}[1]{#1}
\makeatletter
\providecommand{\doi}
  {\begingroup\let\do\@makeother\dospecials
  \catcode`\{=1 \catcode`\}=2 \doi@aux}
\providecommand{\doi@aux}[1]{\endgroup\texttt{#1}}
\makeatother
\providecommand*\mcitethebibliography{\thebibliography}
\csname @ifundefined\endcsname{endmcitethebibliography}
  {\let\endmcitethebibliography\endthebibliography}{}
\begin{mcitethebibliography}{58}
\providecommand*\natexlab[1]{#1}
\providecommand*\mciteSetBstSublistMode[1]{}
\providecommand*\mciteSetBstMaxWidthForm[2]{}
\providecommand*\mciteBstWouldAddEndPuncttrue
  {\def\EndOfBibitem{\unskip.}}
\providecommand*\mciteBstWouldAddEndPunctfalse
  {\let\EndOfBibitem\relax}
\providecommand*\mciteSetBstMidEndSepPunct[3]{}
\providecommand*\mciteSetBstSublistLabelBeginEnd[3]{}
\providecommand*\EndOfBibitem{}
\mciteSetBstSublistMode{f}
\mciteSetBstMaxWidthForm{subitem}{(\alph{mcitesubitemcount})}
\mciteSetBstSublistLabelBeginEnd
  {\mcitemaxwidthsubitemform\space}
  {\relax}
  {\relax}

\bibitem[Ashkin(1970)]{RN2}
Ashkin,~A. Acceleration and trapping of particles by radiation pressure.
  \emph{Phys. Rev. Lett.} \textbf{1970}, \emph{24}, 156--159\relax
\mciteBstWouldAddEndPuncttrue
\mciteSetBstMidEndSepPunct{\mcitedefaultmidpunct}
{\mcitedefaultendpunct}{\mcitedefaultseppunct}\relax
\EndOfBibitem
\bibitem[Ashkin \latin{et~al.}(1986)Ashkin, Dziedzic, Bjorkholm, and
  Chu]{Ashkin1986}
Ashkin,~A.; Dziedzic,~J.~M.; Bjorkholm,~J.~E.; Chu,~S. Observation of a
  single-beam gradient force optical trap for dielectric particles. \emph{Opt.
  Lett.} \textbf{1986}, \emph{11}, 288--290\relax
\mciteBstWouldAddEndPuncttrue
\mciteSetBstMidEndSepPunct{\mcitedefaultmidpunct}
{\mcitedefaultendpunct}{\mcitedefaultseppunct}\relax
\EndOfBibitem
\bibitem[Polimeno \latin{et~al.}(2018)Polimeno, Magazzù, Iatì, Patti, Saija,
  {Esposti Boschi}, Donato, Gucciardi, Jones, Volpe, and Maragò]{Polimeno}
Polimeno,~P.; Magazzù,~A.; Iatì,~M.~A.; Patti,~F.; Saija,~R.; {Esposti
  Boschi},~C.~D.; Donato,~M.~G.; Gucciardi,~P.~G.; Jones,~P.~H.; Volpe,~G.;
  Maragò,~O.~M. Optical tweezers and their applications. \emph{Journal of
  Quantitative Spectroscopy and Radiative Transfer} \textbf{2018}, \emph{218},
  131--150\relax
\mciteBstWouldAddEndPuncttrue
\mciteSetBstMidEndSepPunct{\mcitedefaultmidpunct}
{\mcitedefaultendpunct}{\mcitedefaultseppunct}\relax
\EndOfBibitem
\bibitem[Bouloumis and {Nic Chormaic}(2020)Bouloumis, and {Nic
  Chormaic}]{Bouloumis2020FromFT}
Bouloumis,~T.~D.; {Nic Chormaic},~S. From Far-Field to Near-Field Micro- and
  Nanoparticle Optical Trapping. \emph{Applied Sciences} \textbf{2020},
  \emph{10}\relax
\mciteBstWouldAddEndPuncttrue
\mciteSetBstMidEndSepPunct{\mcitedefaultmidpunct}
{\mcitedefaultendpunct}{\mcitedefaultseppunct}\relax
\EndOfBibitem
\bibitem[Wang \latin{et~al.}(2009)Wang, Schonbrun, and Crozier]{Crozier2009}
Wang,~K.; Schonbrun,~E.; Crozier,~K.~B. Propulsion of Gold Nanoparticles with
  Surface Plasmon Polaritons: Evidence of Enhanced Optical Force from
  Near-Field Coupling between Gold Particle and Gold Film. \emph{Nano Letters}
  \textbf{2009}, \emph{9}, 2623--2629\relax
\mciteBstWouldAddEndPuncttrue
\mciteSetBstMidEndSepPunct{\mcitedefaultmidpunct}
{\mcitedefaultendpunct}{\mcitedefaultseppunct}\relax
\EndOfBibitem
\bibitem[Zhang \latin{et~al.}(2010)Zhang, Huang, Santschi, and
  Martin]{Zhang2010}
Zhang,~W.; Huang,~L.; Santschi,~C.; Martin,~O. J.~F. Trapping and Sensing 10 nm
  Metal Nanoparticles Using Plasmonic Dipole Antennas. \emph{Nano Letters}
  \textbf{2010}, \emph{10}, 1006--1011\relax
\mciteBstWouldAddEndPuncttrue
\mciteSetBstMidEndSepPunct{\mcitedefaultmidpunct}
{\mcitedefaultendpunct}{\mcitedefaultseppunct}\relax
\EndOfBibitem
\bibitem[Fränzl and Cichos(2022)Fränzl, and Cichos]{Cichos2022}
Fränzl,~M.; Cichos,~F. Hydrodynamic manipulation of nano-objects by optically
  induced thermo-osmotic flows. \emph{Nature Communications} \textbf{2022},
  \emph{13}, 656\relax
\mciteBstWouldAddEndPuncttrue
\mciteSetBstMidEndSepPunct{\mcitedefaultmidpunct}
{\mcitedefaultendpunct}{\mcitedefaultseppunct}\relax
\EndOfBibitem
\bibitem[Righini \latin{et~al.}(2009)Righini, Ghenuche, Cherukulappurath,
  Myroshnychenko, de~abajo, and Quidant]{Righini2009}
Righini,~M.; Ghenuche,~P.; Cherukulappurath,~S.; Myroshnychenko,~V.;
  de~abajo,~F. J.~G.; Quidant,~R. Nano-optical trapping of Rayleigh particles
  and Escherichia coli bacteria with resonant optical antennas. \emph{Nano
  letters} \textbf{2009}, \emph{9 10}, 3387--91\relax
\mciteBstWouldAddEndPuncttrue
\mciteSetBstMidEndSepPunct{\mcitedefaultmidpunct}
{\mcitedefaultendpunct}{\mcitedefaultseppunct}\relax
\EndOfBibitem
\bibitem[Verschueren \latin{et~al.}(2019)Verschueren, Shi, and
  Dekker]{Verschueren2019}
Verschueren,~D.; Shi,~X.; Dekker,~C. Nano-Optical Tweezing of Single Proteins
  in Plasmonic Nanopores. \emph{Small Methods} \textbf{2019}, \emph{3},
  1800465\relax
\mciteBstWouldAddEndPuncttrue
\mciteSetBstMidEndSepPunct{\mcitedefaultmidpunct}
{\mcitedefaultendpunct}{\mcitedefaultseppunct}\relax
\EndOfBibitem
\bibitem[Peng \latin{et~al.}(2021)Peng, Kotnala, Rajeeva, Wang, Yao, Bhatt,
  Penley, and Zheng]{Peng2021}
Peng,~X.; Kotnala,~A.; Rajeeva,~B.~B.; Wang,~M.; Yao,~K.; Bhatt,~N.;
  Penley,~D.; Zheng,~Y. Plasmonic Nanotweezers and Nanosensors for
  Point-of-Care Applications. \emph{Adv Opt Mater} \textbf{2021},
  \emph{9}\relax
\mciteBstWouldAddEndPuncttrue
\mciteSetBstMidEndSepPunct{\mcitedefaultmidpunct}
{\mcitedefaultendpunct}{\mcitedefaultseppunct}\relax
\EndOfBibitem
\bibitem[Xu and Crozier(2019)Xu, and Crozier]{Xu2019AlldielectricNF}
Xu,~Z.; Crozier,~K.~B. All-dielectric nanotweezers for trapping and observation
  of a single quantum dot. \emph{Optics express} \textbf{2019}, \emph{27 4},
  4034--4045\relax
\mciteBstWouldAddEndPuncttrue
\mciteSetBstMidEndSepPunct{\mcitedefaultmidpunct}
{\mcitedefaultendpunct}{\mcitedefaultseppunct}\relax
\EndOfBibitem
\bibitem[Hong \latin{et~al.}(2021)Hong, Yang, Kravchenko, and
  Ndukaife]{Hong2021ElectrothermoplasmonicTA}
Hong,~C.; Yang,~S.; Kravchenko,~I.~I.; Ndukaife,~J.~C. Electrothermoplasmonic
  Trapping and Dynamic Manipulation of Single Colloidal Nanodiamond. \emph{Nano
  Letters} \textbf{2021}, \emph{21}, 4921--4927\relax
\mciteBstWouldAddEndPuncttrue
\mciteSetBstMidEndSepPunct{\mcitedefaultmidpunct}
{\mcitedefaultendpunct}{\mcitedefaultseppunct}\relax
\EndOfBibitem
\bibitem[Jiang \latin{et~al.}(2021)Jiang, Roy, Claude, and
  Wenger]{Jiang2021SinglePS}
Jiang,~Q.; Roy,~P.; Claude,~J.-B.; Wenger,~J. Single Photon Source from a
  Nanoantenna-Trapped Single Quantum Dot. \emph{Nano Letters} \textbf{2021},
  \emph{21}, 7030--7036\relax
\mciteBstWouldAddEndPuncttrue
\mciteSetBstMidEndSepPunct{\mcitedefaultmidpunct}
{\mcitedefaultendpunct}{\mcitedefaultseppunct}\relax
\EndOfBibitem
\bibitem[Roxworthy \latin{et~al.}(2012)Roxworthy, Ko, Kumar, Fung, Chow, Liu,
  Fang, and Toussaint]{Roxworthy2012ApplicationOP}
Roxworthy,~B.~J.; Ko,~K.~D.; Kumar,~A.; Fung,~K.~H.; Chow,~E. K.~C.;
  Liu,~G.~L.; Fang,~N.~X.; Toussaint,~K.~C. Application of plasmonic bowtie
  nanoantenna arrays for optical trapping, stacking, and sorting. \emph{Nano
  letters} \textbf{2012}, \emph{12 2}, 796--801\relax
\mciteBstWouldAddEndPuncttrue
\mciteSetBstMidEndSepPunct{\mcitedefaultmidpunct}
{\mcitedefaultendpunct}{\mcitedefaultseppunct}\relax
\EndOfBibitem
\bibitem[Jiang \latin{et~al.}(2020)Jiang, Rogez, Claude, Baffou, and
  Wenger]{Jiang2020QuantifyingTR}
Jiang,~Q.; Rogez,~B.; Claude,~J.-B.; Baffou,~G.; Wenger,~J. Quantifying the
  Role of the Surfactant and the Thermophoretic Force in Plasmonic Nano-optical
  Trapping. \emph{Nano letters} \textbf{2020}, \relax
\mciteBstWouldAddEndPunctfalse
\mciteSetBstMidEndSepPunct{\mcitedefaultmidpunct}
{}{\mcitedefaultseppunct}\relax
\EndOfBibitem
\bibitem[Kotsifaki and {Nic Chormaic}(2022)Kotsifaki, and {Nic
  Chormaic}]{Kotsifaki2022TheRO}
Kotsifaki,~D.~G.; {Nic Chormaic},~S. The role of temperature-induced effects
  generated by plasmonic nanostructures on particle delivery and manipulation:
  a review. \emph{Nanophotonics} \textbf{2022}, \emph{11}, 2199 -- 2218\relax
\mciteBstWouldAddEndPuncttrue
\mciteSetBstMidEndSepPunct{\mcitedefaultmidpunct}
{\mcitedefaultendpunct}{\mcitedefaultseppunct}\relax
\EndOfBibitem
\bibitem[Juan \latin{et~al.}(2011)Juan, Righini, and
  Quidant]{Juan2011PlasmonNT}
Juan,~M.~L.; Righini,~M.; Quidant,~R. Plasmon nano-optical tweezers.
  \emph{Nature Photonics} \textbf{2011}, \emph{5}, 349--356\relax
\mciteBstWouldAddEndPuncttrue
\mciteSetBstMidEndSepPunct{\mcitedefaultmidpunct}
{\mcitedefaultendpunct}{\mcitedefaultseppunct}\relax
\EndOfBibitem
\bibitem[Saleh and Dionne(2012)Saleh, and Dionne]{Saleh2012TowardEO}
Saleh,~A. A.~E.; Dionne,~J.~A. Toward efficient optical trapping of sub-10-nm
  particles with coaxial plasmonic apertures. \emph{Nano letters}
  \textbf{2012}, \emph{12 11}, 5581--6\relax
\mciteBstWouldAddEndPuncttrue
\mciteSetBstMidEndSepPunct{\mcitedefaultmidpunct}
{\mcitedefaultendpunct}{\mcitedefaultseppunct}\relax
\EndOfBibitem
\bibitem[Kotsifaki and {Nic Chormaic}(2019)Kotsifaki, and {Nic
  Chormaic}]{Kotsifaki2019PlasmonicOT}
Kotsifaki,~D.~G.; {Nic Chormaic},~S. Plasmonic optical tweezers based on
  nanostructures: fundamentals, advances and prospects. \emph{Nanophotonics}
  \textbf{2019}, \emph{8}, 1227 -- 1245\relax
\mciteBstWouldAddEndPuncttrue
\mciteSetBstMidEndSepPunct{\mcitedefaultmidpunct}
{\mcitedefaultendpunct}{\mcitedefaultseppunct}\relax
\EndOfBibitem
\bibitem[Bouloumis \latin{et~al.}(2020)Bouloumis, Kotsifaki, Han, {Nic
  Chormaic}, and Truong]{Bouloumis2020FastAE}
Bouloumis,~T.; Kotsifaki,~D.~G.; Han,~X.; {Nic Chormaic},~S.; Truong,~V.~G.
  Fast and efficient nanoparticle trapping using plasmonic connected nanoring
  apertures. \emph{Nanotechnology} \textbf{2020}, \emph{32}, 025507\relax
\mciteBstWouldAddEndPuncttrue
\mciteSetBstMidEndSepPunct{\mcitedefaultmidpunct}
{\mcitedefaultendpunct}{\mcitedefaultseppunct}\relax
\EndOfBibitem
\bibitem[Juan \latin{et~al.}(2009)Juan, Gordon, Pang, Eftekhari, and
  Quidant]{Juan2009SelfB}
Juan,~M.~L.; Gordon,~R.; Pang,~Y.; Eftekhari,~F.; Quidant,~R. Self -induced
  back-action optical trapping of dielectric nanoparticles. \emph{Nature
  Physics} \textbf{2009}, \emph{5}, 915--919\relax
\mciteBstWouldAddEndPuncttrue
\mciteSetBstMidEndSepPunct{\mcitedefaultmidpunct}
{\mcitedefaultendpunct}{\mcitedefaultseppunct}\relax
\EndOfBibitem
\bibitem[Neumeier \latin{et~al.}(2015)Neumeier, Quidant, and
  Chang]{Neumeier2015SelfinducedBO}
Neumeier,~L.; Quidant,~R.; Chang,~D.~E. Self-induced back-action optical
  trapping in nanophotonic systems. \emph{New Journal of Physics}
  \textbf{2015}, \emph{17}, 123008\relax
\mciteBstWouldAddEndPuncttrue
\mciteSetBstMidEndSepPunct{\mcitedefaultmidpunct}
{\mcitedefaultendpunct}{\mcitedefaultseppunct}\relax
\EndOfBibitem
\bibitem[Descharmes \latin{et~al.}(2013)Descharmes, Dharanipathy, Diao, Tonin,
  and Houdr{\'e}]{Descharmes2013ObservationOB}
Descharmes,~N.; Dharanipathy,~U.~P.; Diao,~Z.; Tonin,~M.; Houdr{\'e},~R.
  Observation of backaction and self-induced trapping in a planar hollow
  photonic crystal cavity. \emph{Physical review letters} \textbf{2013},
  \emph{110 12}, 123601\relax
\mciteBstWouldAddEndPuncttrue
\mciteSetBstMidEndSepPunct{\mcitedefaultmidpunct}
{\mcitedefaultendpunct}{\mcitedefaultseppunct}\relax
\EndOfBibitem
\bibitem[Mestres \latin{et~al.}(2016)Mestres, Berthelot, Acimovic, and
  Quidant]{Mestres2016Unravelling}
Mestres,~P.; Berthelot,~J.; Acimovic,~S.~S.; Quidant,~R. Unraveling the
  optomechanical nature of plasmonic trapping. \emph{Light. Sci. Appl.}
  \textbf{2016}, \emph{5}, e16092\relax
\mciteBstWouldAddEndPuncttrue
\mciteSetBstMidEndSepPunct{\mcitedefaultmidpunct}
{\mcitedefaultendpunct}{\mcitedefaultseppunct}\relax
\EndOfBibitem
\bibitem[Zhang \latin{et~al.}(2019)Zhang, Li, Park, feng Su, Goddard, and
  Gelfand]{Zhang2019OptimizationOM}
Zhang,~C.; Li,~J.; Park,~J.~G.; feng Su,~Y.; Goddard,~R.~E.; Gelfand,~R.~M.
  Optimization of metallic nanoapertures at short-wave infrared wavelengths for
  self-induced back-action trapping. \emph{Applied optics} \textbf{2019},
  \emph{58 35}, 9498--9504\relax
\mciteBstWouldAddEndPuncttrue
\mciteSetBstMidEndSepPunct{\mcitedefaultmidpunct}
{\mcitedefaultendpunct}{\mcitedefaultseppunct}\relax
\EndOfBibitem
\bibitem[Zhu \latin{et~al.}(2018)Zhu, Cao, Wang, Nie, Cao, Sun, Jiang,
  Nieto-Vesperinas, Liu, Qiu, and Ding]{Zhu2018}
Zhu,~T.; Cao,~Y.; Wang,~L.; Nie,~Z.; Cao,~T.; Sun,~F.; Jiang,~Z.;
  Nieto-Vesperinas,~M.; Liu,~Y.; Qiu,~C.-W.; Ding,~W. Self-Induced Backaction
  Optical Pulling Force. \emph{Phys. Rev. Lett.} \textbf{2018}, \emph{120},
  123901\relax
\mciteBstWouldAddEndPuncttrue
\mciteSetBstMidEndSepPunct{\mcitedefaultmidpunct}
{\mcitedefaultendpunct}{\mcitedefaultseppunct}\relax
\EndOfBibitem
\bibitem[Daniel and Astruc(2004)Daniel, and Astruc]{GoldNPsSynthesisReview}
Daniel,~M.-C.; Astruc,~D. Gold Nanoparticles: Assembly, Supramolecular
  Chemistry, Quantum-Size-Related Properties, and Applications toward Biology,
  Catalysis, and Nanotechnology. \emph{Chemical Reviews} \textbf{2004},
  \emph{104}, 293--346\relax
\mciteBstWouldAddEndPuncttrue
\mciteSetBstMidEndSepPunct{\mcitedefaultmidpunct}
{\mcitedefaultendpunct}{\mcitedefaultseppunct}\relax
\EndOfBibitem
\bibitem[Grammatikopoulos \latin{et~al.}(2016)Grammatikopoulos, Steinhauer,
  Vernieres, Singh, and Sowwan]{Grammatikopoulos2016NPs}
Grammatikopoulos,~P.; Steinhauer,~S.; Vernieres,~J.; Singh,~V.; Sowwan,~M.
  Nanoparticle design by gas-phase synthesis. \emph{Advances in Physics: X}
  \textbf{2016}, \emph{1}, 81--100\relax
\mciteBstWouldAddEndPuncttrue
\mciteSetBstMidEndSepPunct{\mcitedefaultmidpunct}
{\mcitedefaultendpunct}{\mcitedefaultseppunct}\relax
\EndOfBibitem
\bibitem[Halas \latin{et~al.}(2011)Halas, Lal, Chang, Link, and
  Nordlander]{HalasNPsreview}
Halas,~N.~J.; Lal,~S.; Chang,~W.-S.; Link,~S.; Nordlander,~P. Plasmons in
  Strongly Coupled Metallic Nanostructures. \emph{Chemical Reviews}
  \textbf{2011}, \emph{111}, 3913--3961\relax
\mciteBstWouldAddEndPuncttrue
\mciteSetBstMidEndSepPunct{\mcitedefaultmidpunct}
{\mcitedefaultendpunct}{\mcitedefaultseppunct}\relax
\EndOfBibitem
\bibitem[Mie(1908)]{Mie1908}
Mie,~G. Beiträge zur Optik trüber Medien, speziell kolloidaler
  Metallösungen. \emph{Annalen der Physik} \textbf{1908}, \emph{330},
  377--445\relax
\mciteBstWouldAddEndPuncttrue
\mciteSetBstMidEndSepPunct{\mcitedefaultmidpunct}
{\mcitedefaultendpunct}{\mcitedefaultseppunct}\relax
\EndOfBibitem
\bibitem[Jeanmaire and Van~Duyne(1977)Jeanmaire, and Van~Duyne]{SERS1978}
Jeanmaire,~D.~L.; Van~Duyne,~R.~P. Surface raman spectroelectrochemistry: Part
  I. Heterocyclic, aromatic, and aliphatic amines adsorbed on the anodized
  silver electrode. \emph{Journal of Electroanalytical Chemistry and
  Interfacial Electrochemistry} \textbf{1977}, \emph{84}, 1--20\relax
\mciteBstWouldAddEndPuncttrue
\mciteSetBstMidEndSepPunct{\mcitedefaultmidpunct}
{\mcitedefaultendpunct}{\mcitedefaultseppunct}\relax
\EndOfBibitem
\bibitem[Notarianni \latin{et~al.}(2014)Notarianni, Vernon, Chou, Aljada, Liu,
  and Motta]{NOTARIANNI201423}
Notarianni,~M.; Vernon,~K.; Chou,~A.; Aljada,~M.; Liu,~J.; Motta,~N. Plasmonic
  effect of gold nanoparticles in organic solar cells. \emph{Solar Energy}
  \textbf{2014}, \emph{106}, 23--37, Third and Fourth Generation Solar
  Cells\relax
\mciteBstWouldAddEndPuncttrue
\mciteSetBstMidEndSepPunct{\mcitedefaultmidpunct}
{\mcitedefaultendpunct}{\mcitedefaultseppunct}\relax
\EndOfBibitem
\bibitem[Nguyen \latin{et~al.}(2018)Nguyen, Kim, Tran, Xu, Kianinia, Toth, and
  Aharonovich]{Nguyen2018NanoassemblyOQ}
Nguyen,~M. A.~P.; Kim,~S.; Tran,~T.~T.; Xu,~Z.; Kianinia,~M.; Toth,~M.;
  Aharonovich,~I. Nanoassembly of quantum emitters in hexagonal boron nitride
  and gold nanospheres. \emph{Nanoscale} \textbf{2018}, \emph{10 5},
  2267--2274\relax
\mciteBstWouldAddEndPuncttrue
\mciteSetBstMidEndSepPunct{\mcitedefaultmidpunct}
{\mcitedefaultendpunct}{\mcitedefaultseppunct}\relax
\EndOfBibitem
\bibitem[Liu and Corma(2018)Liu, and Corma]{Lichen2018}
Liu,~L.; Corma,~A. Metal Catalysts for Heterogeneous Catalysis: From Single
  Atoms to Nanoclusters and Nanoparticles. \emph{Chemical Reviews}
  \textbf{2018}, \emph{118}, 4981--5079\relax
\mciteBstWouldAddEndPuncttrue
\mciteSetBstMidEndSepPunct{\mcitedefaultmidpunct}
{\mcitedefaultendpunct}{\mcitedefaultseppunct}\relax
\EndOfBibitem
\bibitem[Wuenschell \latin{et~al.}(2022)Wuenschell, Jee, Buric, and
  Chorpening]{Wuenschell2022GoldSensing}
Wuenschell,~J.; Jee,~Y.; Buric,~M.; Chorpening,~B. Gold nanoparticle
  incorporated oxide thin films for gas sensing at high temperature. \emph{MRS
  Communications} \textbf{2022}, \emph{12}, 308--314\relax
\mciteBstWouldAddEndPuncttrue
\mciteSetBstMidEndSepPunct{\mcitedefaultmidpunct}
{\mcitedefaultendpunct}{\mcitedefaultseppunct}\relax
\EndOfBibitem
\bibitem[Mirigliano \latin{et~al.}(2021)Mirigliano, Paroli, Martini, Fedrizzi,
  Falqui, Casu, and Milani]{Mirigliano2021Neuromorph}
Mirigliano,~M.; Paroli,~B.; Martini,~G.; Fedrizzi,~M.; Falqui,~A.; Casu,~A.;
  Milani,~P. A binary classifier based on a reconfigurable dense network of
  metallic nanojunctions. \emph{Neuromorphic Computing and Engineering}
  \textbf{2021}, \emph{1}, 024007\relax
\mciteBstWouldAddEndPuncttrue
\mciteSetBstMidEndSepPunct{\mcitedefaultmidpunct}
{\mcitedefaultendpunct}{\mcitedefaultseppunct}\relax
\EndOfBibitem
\bibitem[Ghosh \latin{et~al.}(2008)Ghosh, Han, De, Kim, and
  Rotello]{Ghosh2008GoldNI}
Ghosh,~P.~S.; Han,~G.; De,~M.; Kim,~C.; Rotello,~V.~M. Gold nanoparticles in
  delivery applications. \emph{Advanced drug delivery reviews} \textbf{2008},
  \emph{60 11}, 1307--15\relax
\mciteBstWouldAddEndPuncttrue
\mciteSetBstMidEndSepPunct{\mcitedefaultmidpunct}
{\mcitedefaultendpunct}{\mcitedefaultseppunct}\relax
\EndOfBibitem
\bibitem[Vines \latin{et~al.}(2019)Vines, Yoon, Ryu, Lim, and
  Park]{Vines2019GoldNF}
Vines,~J.~B.; Yoon,~J.-H.; Ryu,~N.-E.; Lim,~D.~J.; Park,~H. Gold Nanoparticles
  for Photothermal Cancer Therapy. \emph{Frontiers in Chemistry} \textbf{2019},
  \emph{7}\relax
\mciteBstWouldAddEndPuncttrue
\mciteSetBstMidEndSepPunct{\mcitedefaultmidpunct}
{\mcitedefaultendpunct}{\mcitedefaultseppunct}\relax
\EndOfBibitem
\bibitem[Carabineiro(2017)]{CarabineiroGoldVaccines}
Carabineiro,~S.~A. Applications of Gold Nanoparticles in Nanomedicine: Recent
  Advances in Vaccines. 2017\relax
\mciteBstWouldAddEndPuncttrue
\mciteSetBstMidEndSepPunct{\mcitedefaultmidpunct}
{\mcitedefaultendpunct}{\mcitedefaultseppunct}\relax
\EndOfBibitem
\bibitem[Daraee \latin{et~al.}(2016)Daraee, Eatemadi, Abbasi, Aval, Kouhi, and
  Akbarzadeh]{Daraee2016ApplicationOG}
Daraee,~H.; Eatemadi,~A.; Abbasi,~E.; Aval,~S.~F.; Kouhi,~M.; Akbarzadeh,~A.
  Application of gold nanoparticles in biomedical and drug delivery.
  \emph{Artificial Cells, Nanomedicine, and Biotechnology} \textbf{2016},
  \emph{44}, 410 -- 422\relax
\mciteBstWouldAddEndPuncttrue
\mciteSetBstMidEndSepPunct{\mcitedefaultmidpunct}
{\mcitedefaultendpunct}{\mcitedefaultseppunct}\relax
\EndOfBibitem
\bibitem[Nejati \latin{et~al.}(2021)Nejati, Dadashpour, Gharibi, Mellatyar, and
  Akbarzadeh]{Nejati2021BiomedicalAO}
Nejati,~K.; Dadashpour,~M.; Gharibi,~T.; Mellatyar,~H.; Akbarzadeh,~A.
  Biomedical Applications of Functionalized Gold Nanoparticles: A Review.
  \emph{Journal of Cluster Science} \textbf{2021}, \emph{33}, 1--16\relax
\mciteBstWouldAddEndPuncttrue
\mciteSetBstMidEndSepPunct{\mcitedefaultmidpunct}
{\mcitedefaultendpunct}{\mcitedefaultseppunct}\relax
\EndOfBibitem
\bibitem[Min \latin{et~al.}(2013)Min, Shen, Shen, Zhang, Fang, Yuan, Du, Zhu,
  Lei, and Yuan]{Min2013PlasmonicTrapping}
Min,~C.; Shen,~Z.; Shen,~J.; Zhang,~Y.; Fang,~H.; Yuan,~G.; Du,~L.; Zhu,~S.;
  Lei,~T.; Yuan,~X. Focused plasmonic trapping of metallic particles.
  \emph{Nature Communications} \textbf{2013}, \emph{4}, 2891\relax
\mciteBstWouldAddEndPuncttrue
\mciteSetBstMidEndSepPunct{\mcitedefaultmidpunct}
{\mcitedefaultendpunct}{\mcitedefaultseppunct}\relax
\EndOfBibitem
\bibitem[Kotsifaki \latin{et~al.}(2020)Kotsifaki, Truong, and {Nic
  Chormaic}]{Kotsifaki2020FanoResonantAM}
Kotsifaki,~D.~G.; Truong,~V.~G.; {Nic Chormaic},~S. Fano-Resonant, Asymmetric,
  Metamaterial-Assisted Tweezers for Single Nanoparticle Trapping. \emph{Nano
  letters} \textbf{2020}, \emph{20}, 3388–3395\relax
\mciteBstWouldAddEndPuncttrue
\mciteSetBstMidEndSepPunct{\mcitedefaultmidpunct}
{\mcitedefaultendpunct}{\mcitedefaultseppunct}\relax
\EndOfBibitem
\bibitem[Fedotov \latin{et~al.}(2007)Fedotov, Rose, Prosvirnin, Papasimakis,
  and Zheludev]{Fedotov2007}
Fedotov,~V.~A.; Rose,~M.; Prosvirnin,~S.~L.; Papasimakis,~N.; Zheludev,~N.~I.
  Sharp Trapped-Mode Resonances in Planar Metamaterials with a Broken
  Structural Symmetry. \emph{Physical Review Letters} \textbf{2007}, \emph{99},
  147401\relax
\mciteBstWouldAddEndPuncttrue
\mciteSetBstMidEndSepPunct{\mcitedefaultmidpunct}
{\mcitedefaultendpunct}{\mcitedefaultseppunct}\relax
\EndOfBibitem
\bibitem[Papasimakis and Zheludev(2009)Papasimakis, and
  Zheludev]{Papasimakis2009}
Papasimakis,~N.; Zheludev,~N.~I. Metamaterial-Induced Transparency:Sharp Fano
  Resonances and Slow Light. \emph{Optics and Photonics News} \textbf{2009},
  \emph{20}, 22--27\relax
\mciteBstWouldAddEndPuncttrue
\mciteSetBstMidEndSepPunct{\mcitedefaultmidpunct}
{\mcitedefaultendpunct}{\mcitedefaultseppunct}\relax
\EndOfBibitem
\bibitem[Tanaka \latin{et~al.}(2010)Tanaka, Plum, Ou, Uchino, and
  Zheludev]{Tanaka2010}
Tanaka,~K.; Plum,~E.; Ou,~J.~Y.; Uchino,~T.; Zheludev,~N.~I. Multifold
  Enhancement of Quantum Dot Luminescence in Plasmonic Metamaterials.
  \emph{Physical Review Letters} \textbf{2010}, \emph{105}, 227403\relax
\mciteBstWouldAddEndPuncttrue
\mciteSetBstMidEndSepPunct{\mcitedefaultmidpunct}
{\mcitedefaultendpunct}{\mcitedefaultseppunct}\relax
\EndOfBibitem
\bibitem[Luk'yanchuk \latin{et~al.}(2010)Luk'yanchuk, Zheludev, Maier, Halas,
  Nordlander, Giessen, and Chong]{Boris}
Luk'yanchuk,~B.; Zheludev,~N.~I.; Maier,~S.~A.; Halas,~N.~J.; Nordlander,~P.;
  Giessen,~H.; Chong,~T.~C. The Fano resonance in plasmonic nanostructures and
  metamaterials. \emph{Nature Materials} \textbf{2010}, \emph{9},
  707--715\relax
\mciteBstWouldAddEndPuncttrue
\mciteSetBstMidEndSepPunct{\mcitedefaultmidpunct}
{\mcitedefaultendpunct}{\mcitedefaultseppunct}\relax
\EndOfBibitem
\bibitem[Thijssen \latin{et~al.}(2015)Thijssen, Kippenberg, Polman, and
  Verhagen]{Thijssen2015Plasmomechanical}
Thijssen,~R.; Kippenberg,~T.~J.; Polman,~A.; Verhagen,~E. Plasmomechanical
  Resonators Based on Dimer Nanoantennas. \emph{Nano Letters} \textbf{2015},
  \emph{15}, 3971--3976\relax
\mciteBstWouldAddEndPuncttrue
\mciteSetBstMidEndSepPunct{\mcitedefaultmidpunct}
{\mcitedefaultendpunct}{\mcitedefaultseppunct}\relax
\EndOfBibitem
\bibitem[Simmons \latin{et~al.}(1996)Simmons, Finer, Chu, and
  Spudich]{SIMMONS19961813}
Simmons,~R.; Finer,~J.; Chu,~S.; Spudich,~J. Quantitative measurements of force
  and displacement using an optical trap. \emph{Biophysical Journal}
  \textbf{1996}, \emph{70}, 1813--1822\relax
\mciteBstWouldAddEndPuncttrue
\mciteSetBstMidEndSepPunct{\mcitedefaultmidpunct}
{\mcitedefaultendpunct}{\mcitedefaultseppunct}\relax
\EndOfBibitem
\bibitem[Braibanti \latin{et~al.}(2008)Braibanti, Vigolo, and
  Piazza]{Braibanti}
Braibanti,~M.; Vigolo,~D.; Piazza,~R. Does Thermophoretic Mobility Depend on
  Particle Size? \emph{Phys. Rev. Lett.} \textbf{2008}, \emph{100},
  108303\relax
\mciteBstWouldAddEndPuncttrue
\mciteSetBstMidEndSepPunct{\mcitedefaultmidpunct}
{\mcitedefaultendpunct}{\mcitedefaultseppunct}\relax
\EndOfBibitem
\bibitem[Würger(2010)]{WurgerSoret}
Würger,~A. Thermal non-equilibrium transport in colloids. \emph{Reports on
  Progress in Physics} \textbf{2010}, \emph{73}, 126601\relax
\mciteBstWouldAddEndPuncttrue
\mciteSetBstMidEndSepPunct{\mcitedefaultmidpunct}
{\mcitedefaultendpunct}{\mcitedefaultseppunct}\relax
\EndOfBibitem
\bibitem[Rodr{\'i}guez-Sevilla \latin{et~al.}(2018)Rodr{\'i}guez-Sevilla,
  Prorok, Bednarkiewicz, Marqu{\'e}s, Garc{\'i}a-Mart{\'i}n, Sol{\'e},
  Haro‐Gonz{\'a}lez, and Jaque]{RodrguezSevilla2018OpticalFA}
Rodr{\'i}guez-Sevilla,~P.; Prorok,~K.; Bednarkiewicz,~A.; Marqu{\'e}s,~M.~I.;
  Garc{\'i}a-Mart{\'i}n,~A.; Sol{\'e},~J. A.~G.; Haro‐Gonz{\'a}lez,~P.;
  Jaque,~D. Optical Forces at the Nanoscale: Size and Electrostatic Effects.
  \emph{Nano letters} \textbf{2018}, \emph{18 1}, 602--609\relax
\mciteBstWouldAddEndPuncttrue
\mciteSetBstMidEndSepPunct{\mcitedefaultmidpunct}
{\mcitedefaultendpunct}{\mcitedefaultseppunct}\relax
\EndOfBibitem
\bibitem[Griffiths(2017)]{griffiths_2017}
Griffiths,~D.~J. \emph{Introduction to Electrodynamics}, 4th ed.; Cambridge
  University Press, 2017\relax
\mciteBstWouldAddEndPuncttrue
\mciteSetBstMidEndSepPunct{\mcitedefaultmidpunct}
{\mcitedefaultendpunct}{\mcitedefaultseppunct}\relax
\EndOfBibitem
\bibitem[Ko \latin{et~al.}(2007)Ko, Park, Kim, and Kim]{SuKo_Ga}
Ko,~D.-S.; Park,~Y.~M.; Kim,~S.-D.; Kim,~Y.-W. Effective removal of Ga residue
  from focused ion beam using a plasma cleaner. \emph{Ultramicroscopy}
  \textbf{2007}, \emph{107}, 368--373\relax
\mciteBstWouldAddEndPuncttrue
\mciteSetBstMidEndSepPunct{\mcitedefaultmidpunct}
{\mcitedefaultendpunct}{\mcitedefaultseppunct}\relax
\EndOfBibitem
\bibitem[Fan \latin{et~al.}(2018)Fan, Zhang, Shen, Fu, Wei, qiang Li, and
  Soukoulis]{Fan2018AchievingAH}
Fan,~Y.; Zhang,~F.; Shen,~N.; Fu,~Q.; Wei,~Z.; qiang Li,~H.; Soukoulis,~C.~M.
  Achieving a high- Q response in metamaterials by manipulating the toroidal
  excitations. \emph{Physical Review A} \textbf{2018}, \emph{97}, 033816\relax
\mciteBstWouldAddEndPuncttrue
\mciteSetBstMidEndSepPunct{\mcitedefaultmidpunct}
{\mcitedefaultendpunct}{\mcitedefaultseppunct}\relax
\EndOfBibitem
\bibitem[Jones \latin{et~al.}(2015)Jones, Maragò, and
  Volpe]{jones_marago_volpe_2015}
Jones,~P.~H.; Maragò,~O.~M.; Volpe,~G. \emph{Optical Tweezers: Principles and
  Applications}; Cambridge University Press, 2015\relax
\mciteBstWouldAddEndPuncttrue
\mciteSetBstMidEndSepPunct{\mcitedefaultmidpunct}
{\mcitedefaultendpunct}{\mcitedefaultseppunct}\relax
\EndOfBibitem
\bibitem[Rosenblatt \latin{et~al.}(2020)Rosenblatt, Simkhovich, Bartal, and
  Orenstein]{PhysRevX.10.011071}
Rosenblatt,~G.; Simkhovich,~B.; Bartal,~G.; Orenstein,~M. Nonmodal Plasmonics:
  Controlling the Forced Optical Response of Nanostructures. \emph{Phys. Rev.
  X} \textbf{2020}, \emph{10}, 011071\relax
\mciteBstWouldAddEndPuncttrue
\mciteSetBstMidEndSepPunct{\mcitedefaultmidpunct}
{\mcitedefaultendpunct}{\mcitedefaultseppunct}\relax
\EndOfBibitem
\end{mcitethebibliography}

\end{document}